\documentclass[11pt]{elsarticle}

\usepackage{amssymb}

\usepackage{color}
\usepackage{lineno}
\usepackage{hyperref}
\usepackage{ulem}
\usepackage{soul}

\journal{Nuclear Instruments and Methods A}

\begin{document}

\begin{frontmatter}



\title{The STAR Forward Silicon Tracker}



\author[bnl,osu]{J.~D.~Brandenburg}
\author[ncku,purdue]{Y.~Chang}
\author[sdu]{J.~Dong}
\author[sdu]{Y.~He}
\author[bnl,fudan,lbl]{Y.~Hu}
\author[uic]{B.~Huang}
\author[ncku]{H.~Huang}
\author[uic,ncku]{T.~Huang}
\author[ncku,purdue]{H.~Li}
\author[sdu]{M.~Nie}
\author[bnl] {R.~Sharma}
\author[imp,uic]{X.~Sun}
\author[bnl]{P.~Tribedy}
\author[bnl]{F.~Videb\ae k}
\author[ceem]{G.~Visser}
\author[uic]{G.~Wilks}
\author[ncku]{P.~Wang}
\author[uic,ucas]{G.~Xie}
\author[sdu]{G.~Yan}
\author[uic,lbl]{Z.~Ye}
\author[sdu]{L.~Yi}
\author[asiop,ncku]{Y.~Yang}
\author[uic,cqu]{S.~Zhang}
\author[uic]{Z.~Zhang}

\affiliation[bnl]{organization={Brookhaven National Lab},
                 city={Upton},
                 postcode={11973}, 
                 state={New York}
                 }
           
\affiliation[cqu]{organization={Chongqing University},  
                 city={Chongqing},
                 potcode={401331},
                 state={Chongqing}
                 }           
           
\affiliation[fudan]{organization={Fudan University},
                   city={Shanghai},
                   postcode={200433},
                   state={Shanghai}
                   }
           
\affiliation[ceem]{organization={Indiana University}, 
                  city={Bloomington}, 
                  postcode={47408},
                  state={Indiana}
                  }
                  
\affiliation[imp]{organization={Institute of Modern Physics, Chinese Academy of Sciences},
                 city={Lanzhou},
                 postcode={730000}, 
                 state={Gansu}
                 }
\affiliation[asiop]{organization={Institute of Physics, Academia Sinica},                   
                  city={Taipei},
                  postcode={115} 
                  }
                  
\affiliation[lbl]{organization={Lawrence Berkeley National Laboratory},
                 city={Berkeley},
                 postcode={94720}, 
                 state={California}
                 }

\affiliation[ncku]{organization={National Cheng Kung University},                   
                  city={Tainan},
                  postcode={70101} 
                  }

\affiliation[osu]{organization={Ohio State University},  
                 city={Columbus},
                 postcode={43210},
                 state={Ohio}
                 }
        
\affiliation[purdue]{organization={Purdue University},  
                    city={West Lafayette},
                    postcode={47907}, 
                    state={Indiana}
                    }
                 
\affiliation[sdu]{organization={Shandong University},
                city={Qingdao},
                postcode={266237}, 
                state={Shandong}
                }

\affiliation[ucas]{organization={University of Chinese Academy of Sciences},
                 city={Beijing},
                 postcode={100049}, 
                 state={Beijing}
                 }

\affiliation[uic]{organization={University of Illinois at Chicago},
                 city={Chicago},
                 postcode={60607}, 
                 state={Illinois}
                 }

\begin{abstract}
The Forward Silicon Tracker (FST) is a pivotal component of the forward upgrade of the Solenoidal Tracker at RHIC (STAR), designed to discern hadron charge signs with a momentum resolution better than 30\% for $0.2 < p_T < 2$ GeV/c in the $2.5 < \eta < 4$ pseudorapidity range. Its compact design features three disks along the beam direction, minimized material budget, and scattering effects. The FST uses Hamamatsu's p-in-n silicon strip sensors with a double metal layer that enables efficient signal routing to the readout electronics, enhancing overall detector performance. The flexible hybrid boards, essential for the readout system, are constructed with Kapton and copper layers to optimize signal handling and power distribution. These boards connect silicon strips to analogue pipeline ASIC APV25-S1 chips, which read up to 128 channels each. A cooling system with nonconducting, volatile NOVEC 7200 coolant at 22.2°C mitigates ASIC-generated heat. The FST enhances forward tracking performance at STAR as an integral part of the forward upgrade.
\end{abstract}




\end{frontmatter}




\section{\label{sec:introduction}Introduction}

The STAR collaboration has undertaken a significant detector upgrade to expand its forward acceptance, aiming to address several outstanding questions in cold and hot Quantum Chromodynamics (QCD)~\cite{star0648} through the study of $p+p$, $p+$Au and Au+Au collisions at the Relativistic Heavy Ion Collider (RHIC). The forward upgrade program first proposed by STAR in 2018 \cite{fstar}, was approved, researched, constructed, installed, and fully commissioned by the year 2022. One key component of the forward upgrade is a Forward Tracking System (FTS), which was installed and commissioned in 2022. 

The STAR FTS consists of the Forward Silicon Tracker (FST) and the small-strip Thin Gap Chambers (sTGC)~\cite{Shi:2020gex}. 
Both tracking systems are needed to achieve the physics tracking goals of the forward upgrade, namely, the capability of discriminating the charge sign of hadrons for transverse asymmetry studies and separating electrons from photons for Drell-Yan measurements by charge-sign measurements. 
The physics goals for measurements in Au+Au collisions, such as the two-particle correlation measurements, further require a momentum resolution better than 30\% for tracks with 0.2 $<$ $p_T$ $<$ 2 GeV/$c$ and an efficiency of better than 80\% for events with approximately 100 tracks per event.

\begin{figure}[htb!]
\centering
\includegraphics[width=0.60\textwidth]{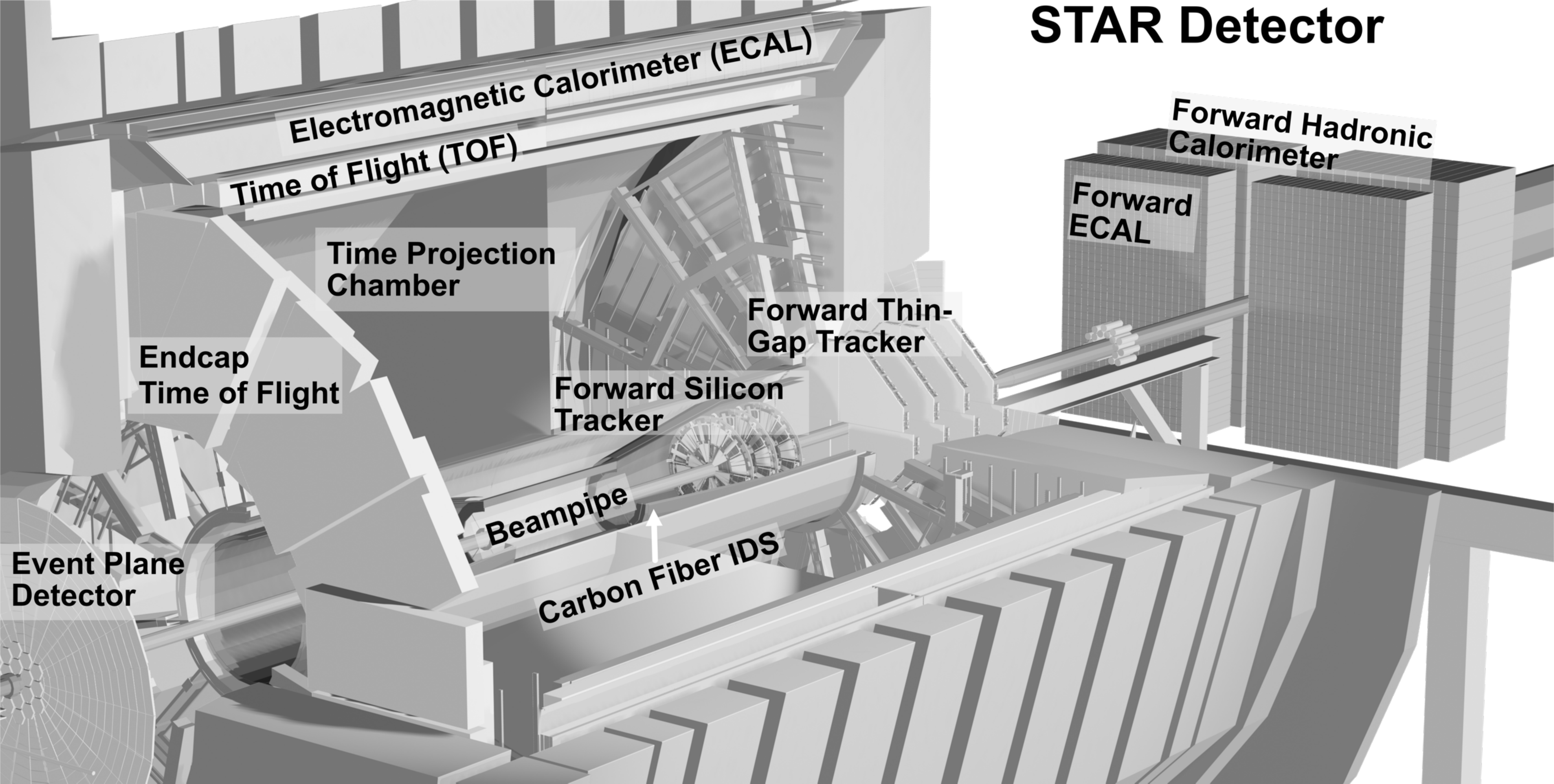}
\includegraphics[width=0.35\textwidth]{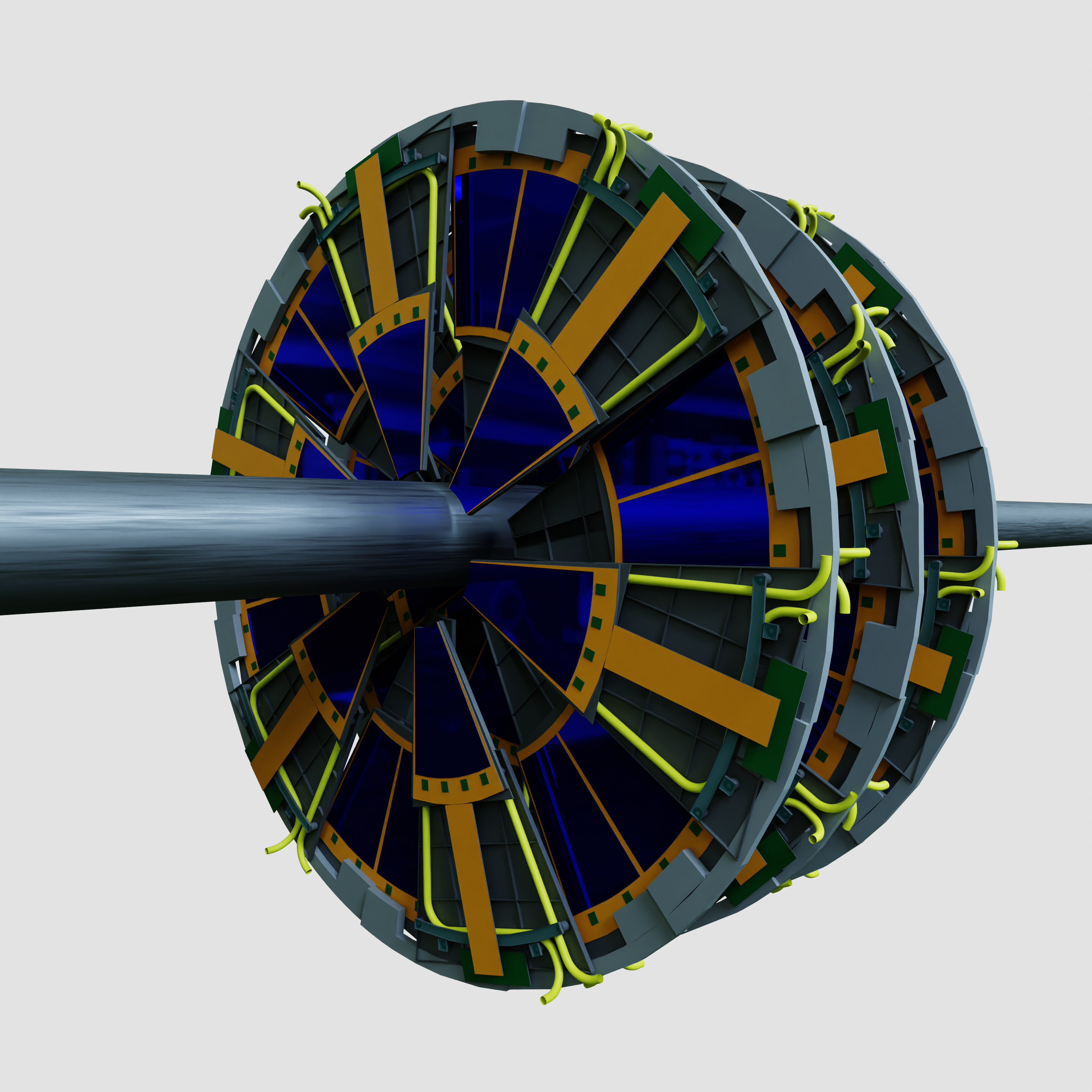}
\caption{Schematic view of STAR detector and forward detector components.}
\label{figs:star}
\end{figure}

The FST consists of 3 disks separated by 13.5 cm oriented orthogonal to the beam line starting at about 152 cm inside the STAR solenoid magnet, which has a magnetic field of 0.5 T,  in the bore of the carbon fiber support cone. A cut-open picture of the STAR magnet is in Figure \ref{figs:star} showing the magnet, TPC cylinder, beam pipe,  the 3 FST disks, and the 4 sTGC planes.
The FST covers polar angles from $1.9^\mathrm{o}$ to $8.9^\mathrm{o}$ with full azimuthal coverage. This corresponds to the pseudo rapidity region of $2.5 < \eta < 4.0$ where $\eta$ is defined as $-\log(\tan(\theta/2))$.

The FST is a compact tracker, featuring segmented silicon strip sensors and APV (Analogue Pipeline Voltage mode)~\cite{FRENCH2001359} front-ends for seamless integration via Kapton-based flexible hybrids. It operates on a minimal material budget ($\sim$1\% $X_0$ per disk) and employs a NOVEC 7200 cooling system to ensure good  performance in RHIC's radiation environment. This article describes the design of the FST, its construction, and its performance.

\section{\label{sec:design}Layout and Design}

The FTS is designed to provide charge-sign separation and photon-electron discrimination. The former requires that the FTS have a high spatial resolution, while the latter calls for a low material budget in the acceptance. 
It is well known that carefully designed Silicon-based detectors can meet such requirements, thanks to their unique properties that are not available with other types of detectors: the combination of extremely precise measurement with high readout speed; direct availability of signals in electronic form; the simultaneous measurement of energy and position; and the possibility of integrating detector and readout electronics on a common substrate.  

\begin{figure}[!ht]
 	\centering
 	\includegraphics[width=0.98\textwidth]{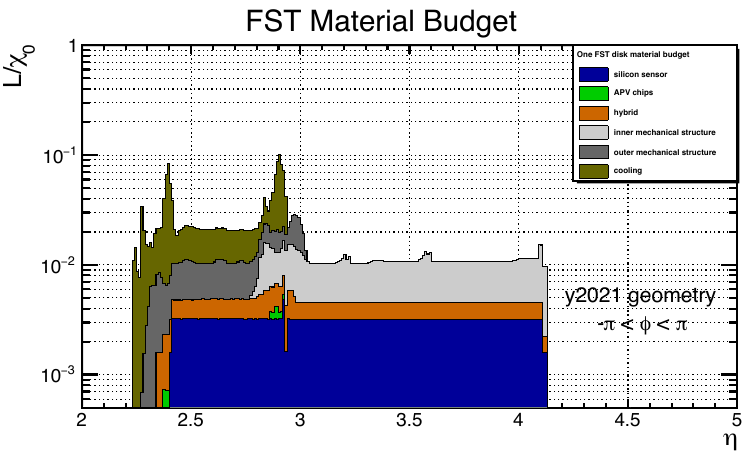}
    \includegraphics[width=0.98\textwidth]{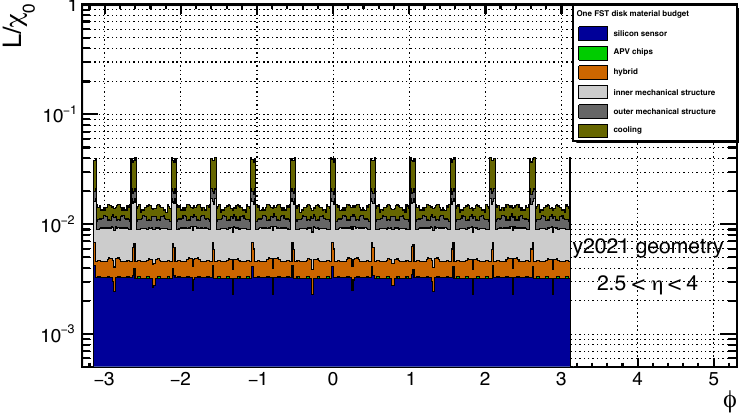}
  
 	\caption{FST material budget in radiation length ($X_0$) as a function of $\eta$ (top) and $\phi$ (bottom) with contributions from different components shown separately.}
 	\label{figs:material}
 \end{figure}

The FST consists of three Silicon detector planes. The layout of the FST studied in simulations is shown in Figure \ref{figs:star}, where three detector planes are located at Z=152, 165, and 179 cm from the nominal interaction point (IP), respectively. Each of the three planes has 12 wedges covering the full 2$\pi$ range in $\phi$ and 2.5-4 in $\eta$. A wedge has 128 ($\phi$) times 8 ($\eta$) strips and provides a $\phi$-resolution of around 1.2 mrad and a $\eta$-resolution of about 0.07. Due to mechanical and size constraints of the silicon sensors, each $30^\mathrm{o}$ wedge consists of 3 sensors, one inner, and two outer sensors. The inner sensor is read out by 4 chips placed near a radius of 16.5 cm, while the 2 outer sensors are read out by 4 front-end readout chips placed near the outer radius edge of 28.0 cm. As shown in Fig.~\ref{figs:material}, the material budget over the most active Silicon sensor area is around 1\% $X_0$, with 0.3\% from the silicon sensor, 0.1\% from the flexible printed circuit board, and 0.6\% from module mechanical structures. The distributed cooling tubes have concentrated local contributions as large as 10\% $X_0$.

The non-ionization energy loss (NIEL) and total irradiation dose (TID) to the FST were estimated based on simulation and operational experience of the previous STAR Silicon detectors. The NIEL and TID were determined to be insignificant (O($10^{12}$) cm$^{-2}$ 1 MeV neutron equivalent flux and O(10) krad, respectively) \cite{fstar}. Therefore, the detector was designed to operate at room temperature. This decision has been confirmed by the observed leakage current change of the FST during its operations with proton-proton collisions in 2022 and Au-Au collisions in 2023 as reported in section \ref{sec:operation}.

 \begin{figure}[!htb]
 	\centering
  	\includegraphics[width=0.9\textwidth]{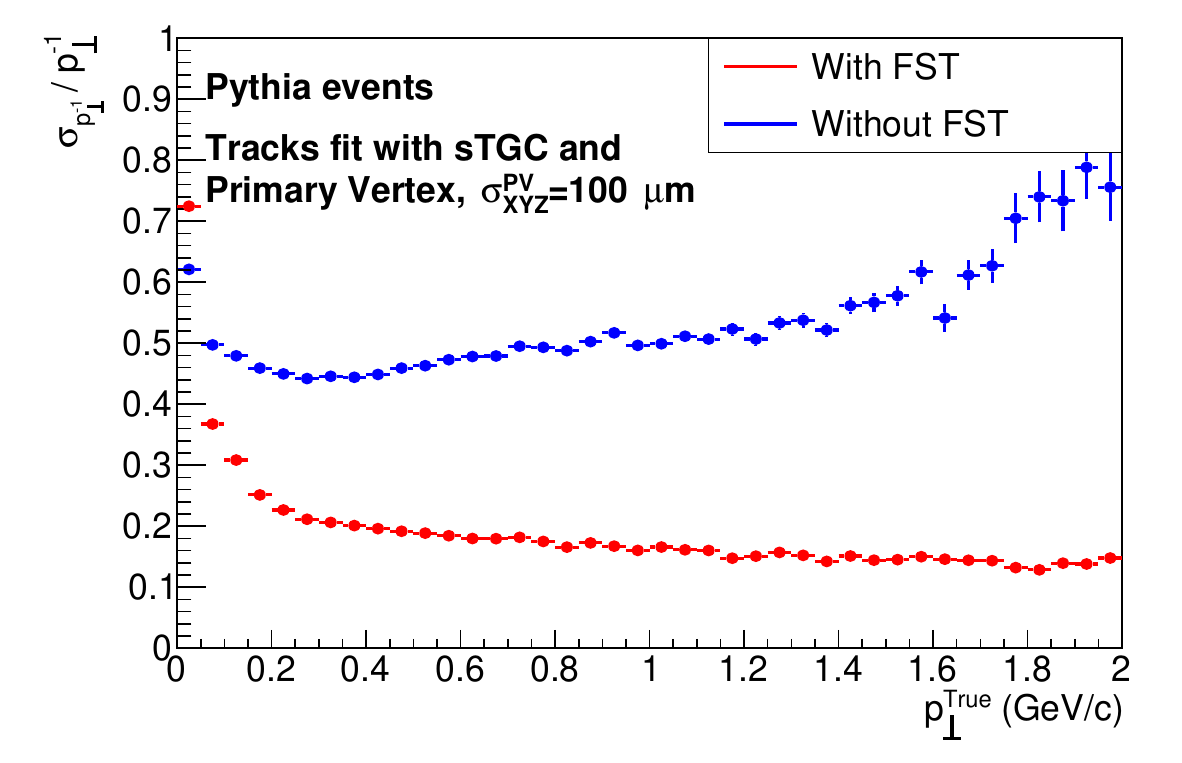}
 	
 	\caption{Resolution of the forward tracking system vs. $p_T$ with and without the FST used in tracking for a sample of PYTHIA p+p 500 GeV/c events in the STAR simulation environment.}
 
 	\label{figs:pythiasim}
 \end{figure}

The performance of the FTS is illustrated in Figure \ref{figs:pythiasim}. A full GEANT3 simulation with Pythia p+p events was carried out to show the expected resolution of curvature $\sigma_{p_T^{-1}}/ p_T^{-1} $ vs. $p_T$.
It shows the improved performance with the addition of the FST tracking stations fulfills the requirement. 

\section{\label{sec:silicon}Silicon Sensors}

The FST Silicon sensors were fabricated by Hamamatsu. A disk has in total 12 sectors, each of which is instrumented with one inner and two outer sensors. The sensor thickness is 320 $\pm$ 15 $\mu$m and the bulk resistivity is around 4$k\Omega\cdot$cm. An inner sensor has 128 ($\phi$) times 4 ($\eta$) strips covering 30 degrees in azimuth and radius from 5.0 to 16.5 cm, while an outer sensor has 64 ($\phi$) times 4 ($\eta$) strips covering 15 degrees in azimuth and radius from 16.5 to 28.0 cm. Each strip covers the same $\phi$ region and has the same length in the radial direction.

\begin{figure}[!ht]
 	\centering
 	\includegraphics[width=0.95\textwidth]{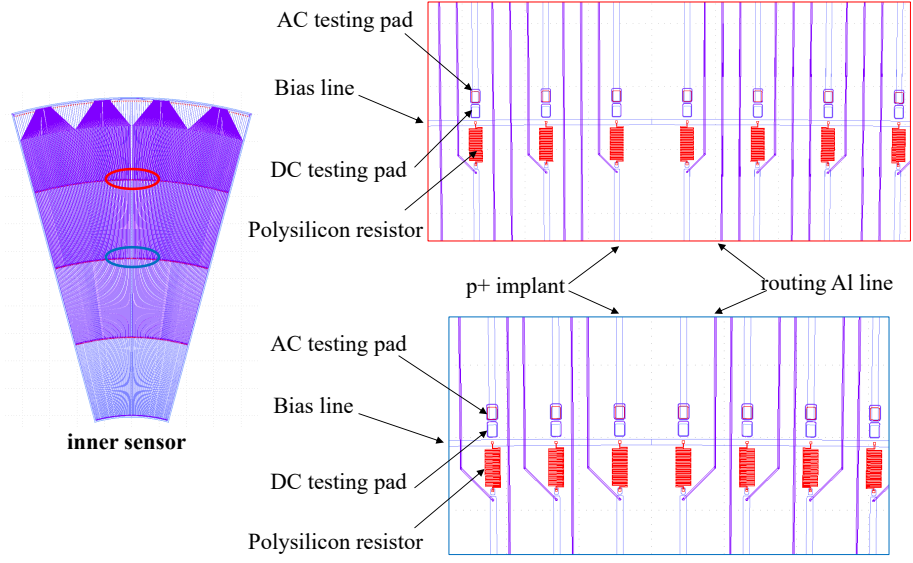}
 	\caption{FST inner Silicon sensor layout. Shown on the right are zoomed in views at the edge of two neighboring strip regions.}
 	\label{figs:sensor1}
 \end{figure}

The Silicon sensors are p-in-n planar sensors with double metal layers. Figure \ref{figs:sensor1} shows the layout of the inner sensor (the outer sensor has a similar layout but half the strips). The top and bottom edges were diced with a polygonal approximation with around 2 mm inactive region on one side. Aluminum strips in the bottom metal layer (not shown) are AC-coupled to p+ implant strips collecting electric signals in the silicon sensor produced by traversing charged particles. Aluminum traces in the top metal layer route the electric signals from the bottom Aluminum strips to the outer edge of the sensor where wire-bonding pads are connected to front-end readout ASICs. Pads at the end of each strip in the top metal layer are connected to the p+ implant (bottom metal) strips and provide direct DC (AC) electric connections to each strip for testing purposes. 

\begin{figure}[!htb]
 	\centering
  	\includegraphics[width=0.9\textwidth]{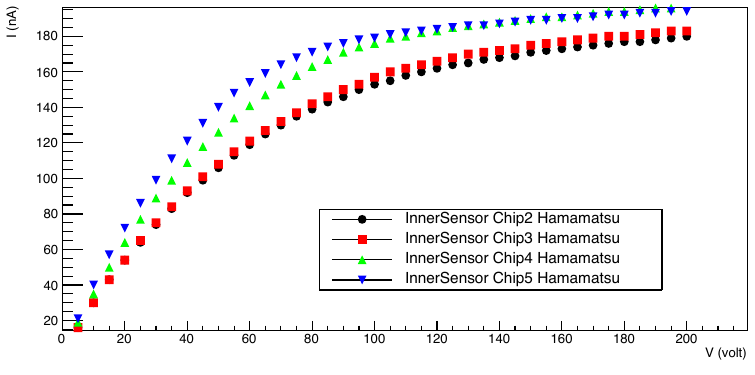}
 	\includegraphics[width=0.9\textwidth]{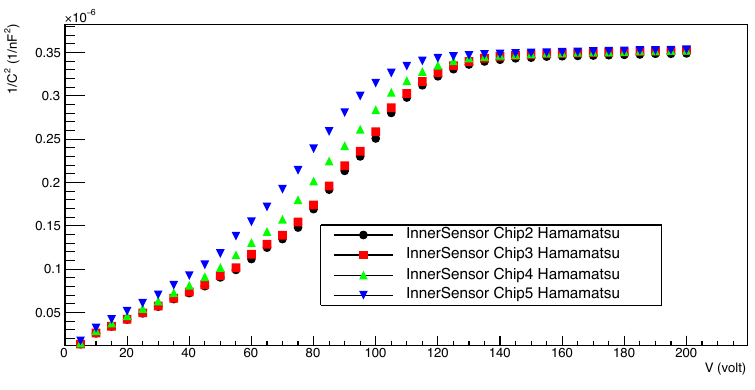}
 	\caption{Leakage current $I$ (top) and inverse capacitance squared $1/C^2$ (bottom) as a function of reverse bias voltage $V$ for inner sensors.}
 	\label{figs:sensor2}
 \end{figure}

A polysilicon resistor connects the p+ implant strip to a p+ bias line (see Figure \ref{figs:sensor1}). The backside of the silicon sensor at a positive electric potential and the p+ bias line at the ground potential provide the needed reverse bias to the silicon sensor. Figure \ref{figs:sensor2} shows the leakage current and inverse capacitance squared as a function of reverse bias voltage for some inner sensors with 100 kHz test frequency. As can be seen, the capacitance reaches a minimum plateau of around 110-130$V$, while the leakage current is below 0.2 $\mu A$ before radiation.  

\section{\label{sec:hybrid}Hybrid}
Flexible substrates hold electronics to form a circuit board known as a "hybrid". To meet the requirement of a low material budget, the flexible hybrid was utilized. The hybrid is one of the key components of the front-end readout system, designed for the placement of silicon strip sensors and analog front-end chips. The transfer board, T-Board, was designed to transfer the hybrid signals and voltages to the readout system.

The silicon sensor has inner and outer sensor parts, thus the hybrid is divided into inner and outer parts accordingly. For inner sensors, 512 silicon strips are designed for signal readout, whereas each outer sensor utilizes 256 silicon strips. To better optimize the geometry, the inner hybrid carries signals from 1 sensor and the outer hybrid carries signals from 2 sensors for each module. Thus 512 readout channels are required for both inner and outer hybrids. 
The APV25-S1 chip~\cite{FRENCH2001359} is chosen to read out the sensor signals as it is an analogue pipeline ASIC intended for the read-out of silicon strip detectors. Each APV25 chip features 128 channels, making 4 APV25 chips sufficient for the inner and outer sensor requirements. To meet the power supply demands of these sensors, we have configured 14 evenly spaced high-voltage lines in the inner hybrid and 16 in the outer hybrid.

\begin{figure}[!ht]
 	\centering
 	\includegraphics[scale=0.16]{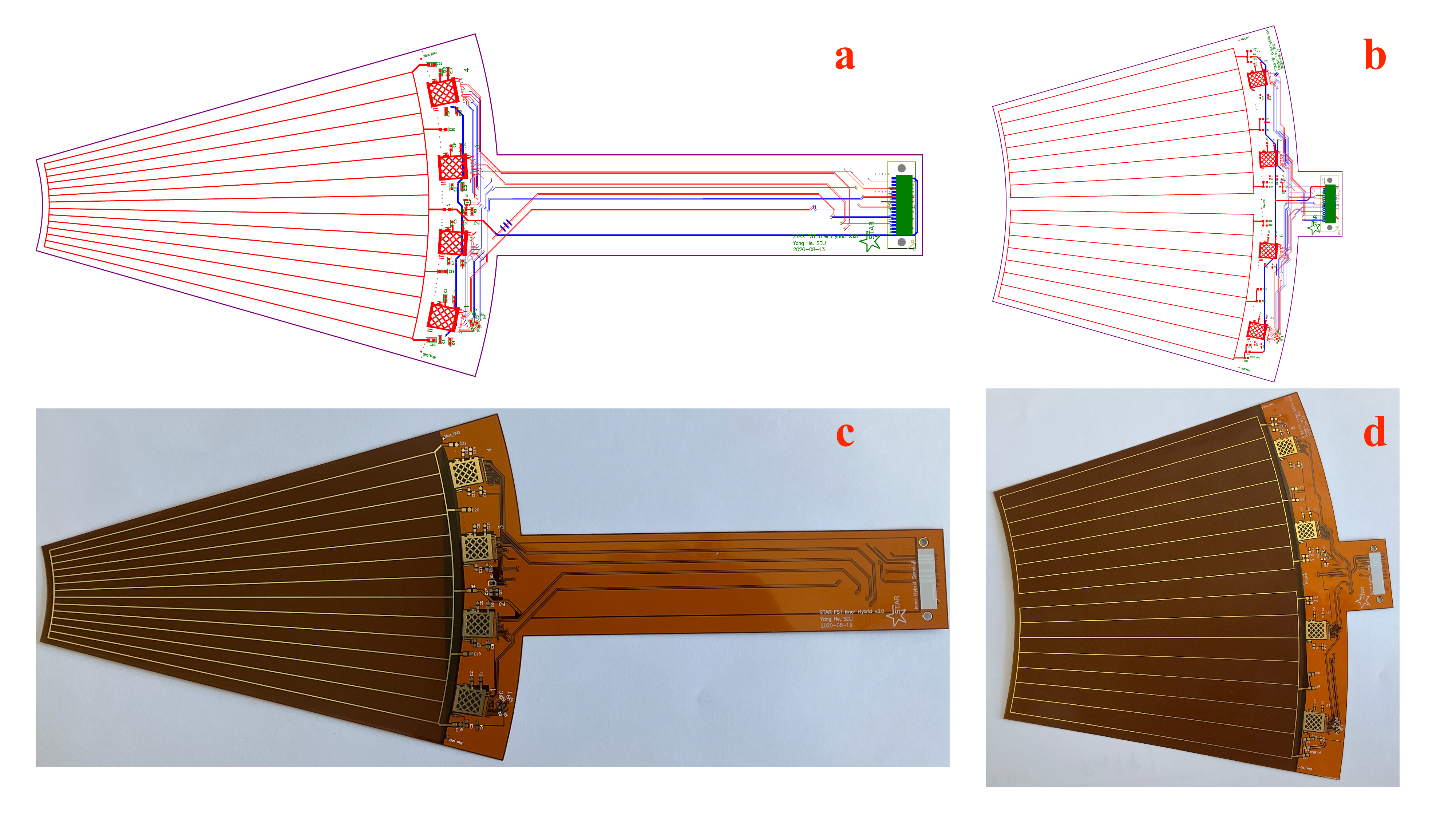}
 	\caption{Designs for the a) inner and b) outer hybrids. Photos of the c) inner and d) outer hybrids.}
 	\label{figs:hybrid}
 \end{figure}

The hybrids are designed by Shandong University. Figure~\ref{figs:hybrid}(a) and (b) show the layouts of inner and outer hybrids, and (c) and (d) are the corresponding photos. The hybrids share the same sector angle of 32 degrees, and the inner and outer hybrids have tail lengths of 12.6 cm and 2cm, respectively. Each hybrid is made of Kapton substrate with 3 copper layers and a thickness of roughly 0.3mm to reduce interactions with passing particles in the sensor area. Both inner and outer hybrids have 4 APV25 chip placement sites, and provide the electrical connection from the APV25 chips to the connector site on the end of each hybrid. The silicon strip sensors and APV25 chips are wire-bonded and glued onto hybrids. Thus, power and signal connectivity from the hybrid to APV25 chips are provided by wire bonds. All hybrids are finished with ENEPIG plating for compatibility with both wire bonding and conventional soldering. 

\section{\label{sec:module}Module Mechanics}
\subsection{Main structure}
\begin{figure}[!ht]
 	\centering
 	\includegraphics[scale=0.9]{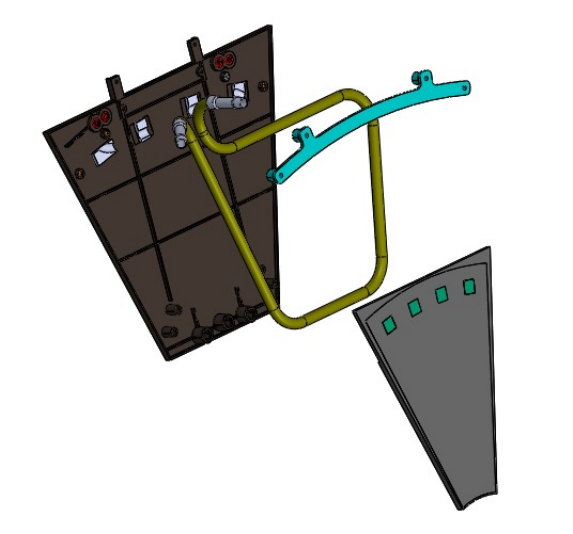}
 	\caption{The main support structure of an FST module of inner and outer wedges, cooling pipe, heat sinks, and tube holder.}
 	\label{figs:module_overview}
 \end{figure}

The main support structure (MS) for each FST module is divided into five major pieces: inner and outer detector wedges, cooling pipe, heat sinks, and the tube holder, as shown in Figure~\ref{figs:module_overview}. The lengths of the inner and outer wedges are 144.5 and 165.4 cm, respectively. The material used for the main support structure is “polyetheretherketone” (PEEK), the cooling pipes are made of 316 stainless steel with a 0.635 cm outer diameter and thickness of 0.5 mm, and the heat sinks are made of Aluminum alloy 6061.

The leakage test on the cooling tube is performed by immersing it in water, injecting 90 psi air for one minute, and then checking for bubbles originating from the connections.
Since the coolant used in the STAR FST is NOVEC 7200, the soft tube for connecting the adjacent wedges is chosen by performing a compatibility test. The test is conducted using the Soxhlet extraction method~\cite{LOPEZBASCON2020327} recommended by 3M. Based on the tests, Tygon 3370 I.B. made by Saint-Gobain is chosen. Since the experimental environment has a high radiation area, the tube is also tested with Co-60 source with the dosage of 3.2 and 16 kGy, where 3.2 kGy is the expected integrated radiation dosage. There are no significant changes observed in this test.

\subsection{Assembly}

\begin{figure}[!ht]
 	      \centering
 	      \includegraphics[scale=0.4]{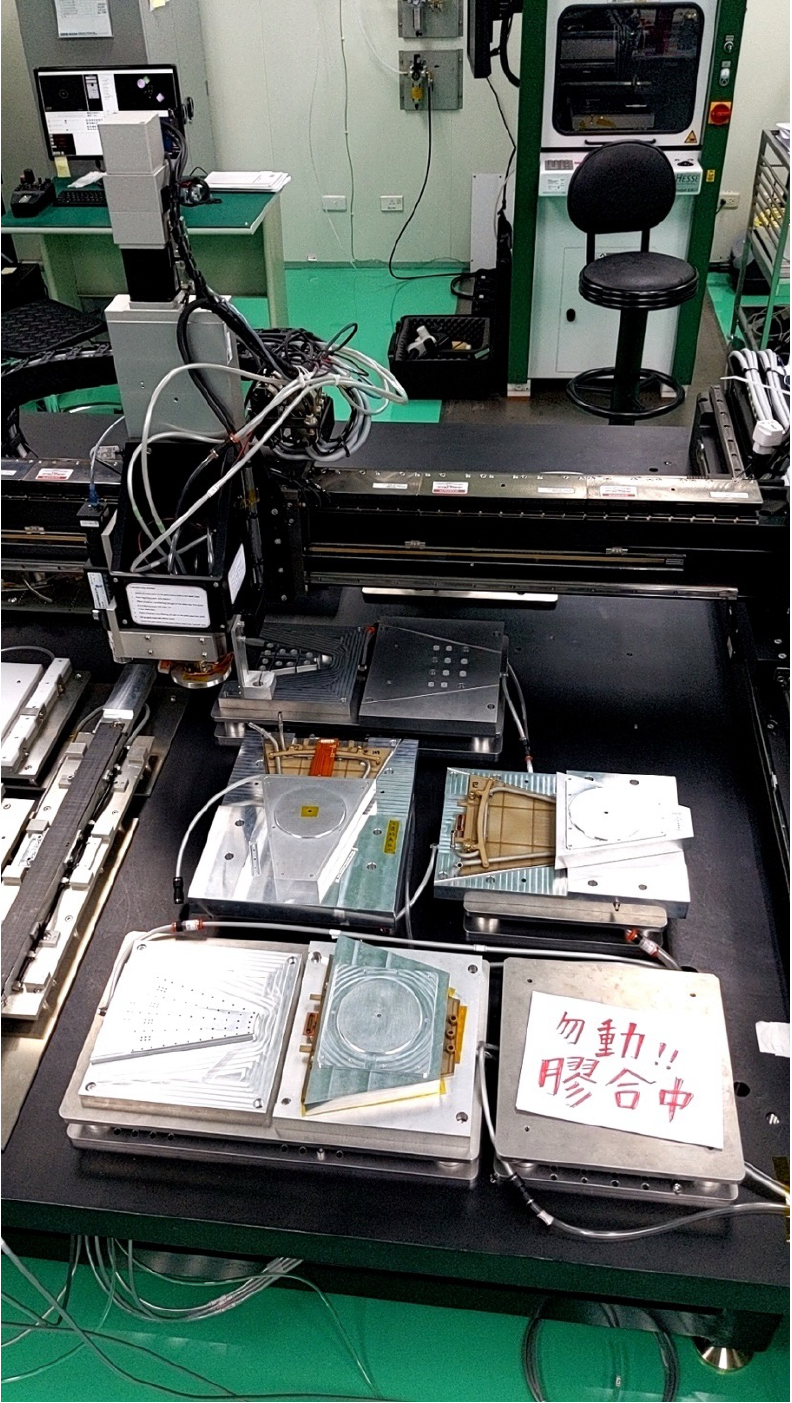}
         	\caption{The STAR FST module assembly at TIDC. }
        	\label{figs:module_assembly_picture}
        \end{figure}

The module assembly is conducted in the assembly Lab at Taiwan Instrumentation Detector Consortium (TIDC)~\footnote{https://www.taiwan-tidc.org}, as shown in Figure~\ref{figs:module_assembly_picture}. The assembly procedure can be divided into 6 parts and they will be described below in detail. 
          
\begin{enumerate}
    \item \textbf{Connector mounting}:
    The Nicomatic connector is soldered on the Flexible Hybrid PCB by an experienced company in Taiwan, then the visual inspection is performed to check the placement and quality of soldering. Finally, the open connection test is performed and the surface is cleaned. 

     \item \textbf{Guide pins Installation}:
     Since the position of the detector module is crucial, the two precision guide pins are designed. To install the guide pins in the correct position, the guide pins are installed by the pin installation jig to make sure the length of the pin exceeds the sensor side of MS is 9.0 mm as shown in Figure~\ref{figs:module_guide_pin_done}.

         \begin{figure}[!ht]
 	      \centering
 	      \includegraphics[scale=0.9]{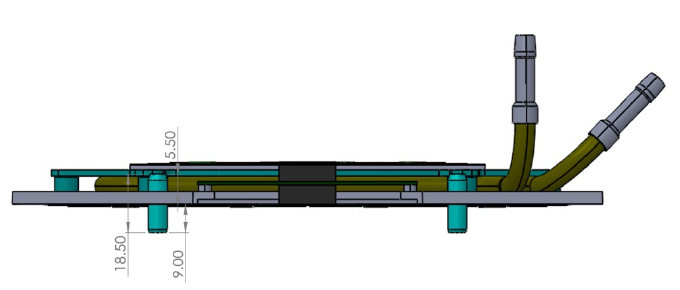}
         	\caption{The positions of guide pins after installation.}
        	\label{figs:module_guide_pin_done}
        \end{figure}

     \item \textbf{Bonding the outer Flexible Hybrid PCB to outer wedge}: 
     To bond the outer Flexible Hybrid PCB to the outer wedge, the first step is to place the pickup tool and two trays for the other wedge and PCB on the assembly table. The outer Flexible Hybrid PCB is picked up by the gantry with a special jig as shown in Figure~\ref{figs:module_robotic_outer}, and the locations are determined by the camera on the gantry using two guide points on the outer Hybrid PCB. Note that ``Araldite 2011''~\cite{Araldite} is used for bonding. 
  
          \begin{figure}[!ht]
           \begin{center}
              \includegraphics[width=0.38\textwidth]{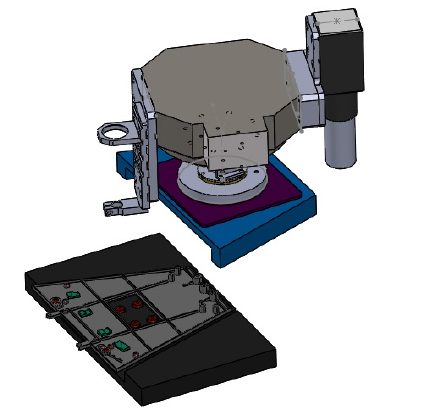}
               \includegraphics[width=0.38\textwidth]{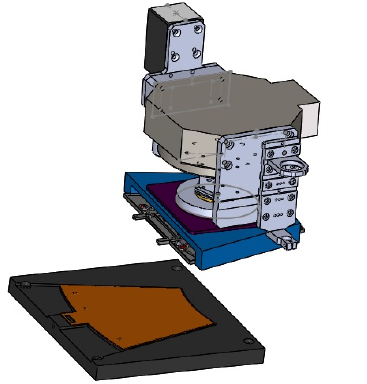}  
             \end{center}
         \caption{ The gantry picks up the outer wedge and places it onto the Hybrid PCB.
           \label{figs:module_robotic_outer}}
          \end{figure}

    \item \textbf{Assembling the inner wedge to the outer wedge}: 
       Before assembling the inner and outer wedge together, the cooling tube and tube holder need to be installed on the outer wedge. Then, the inner and the outer wedge (with Hybrid PCB and cooling tube assembled) are placed on the jig. The robotic gantry picks up the inner wedge and assembles it onto the outer one, as shown in Figure~\ref{figs:module_robotic_inner}.

          \begin{figure}[!ht]
           \begin{center}
              \includegraphics[width=0.3\textwidth]{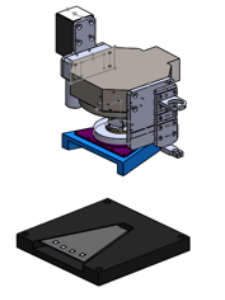}
               \includegraphics[width=0.3\textwidth]{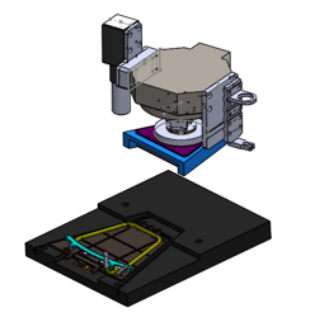}  
               \includegraphics[width=0.3\textwidth]{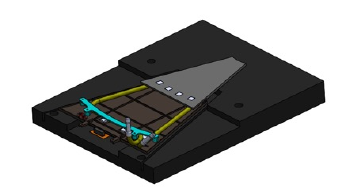}  
             \end{center}
         \caption{ The robotic gantry picks up the inner wedge and assembles it onto the outer wedge. 
           \label{figs:module_robotic_inner}}
          \end{figure}

    \item \textbf{Bonding Inner Hybrid PCB on the inner wedge}:
        Similarly to step 4, the inner Hybrid PCB is bonded on the inner wedge with the gantry. However, in this step, the Hybrid PCB is picked up and glued on the inner wedge, instead of the outer wedge, as shown in Figure~\ref{figs:module_robotic_inner_hybrid}.
        
          \begin{figure}[!ht]
           \begin{center}
              \includegraphics[width=0.3\textwidth]{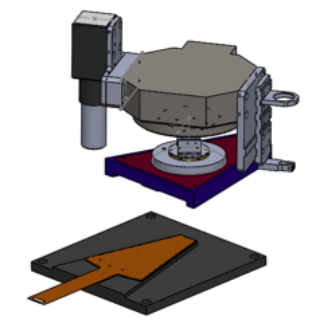}
               \includegraphics[width=0.3\textwidth]{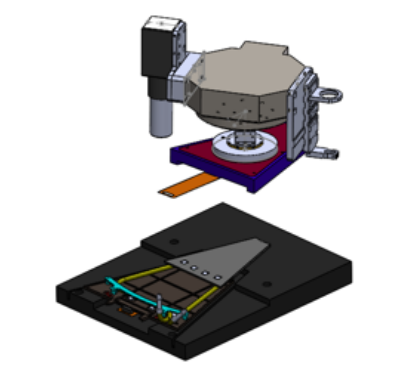}  
               \includegraphics[width=0.3\textwidth]{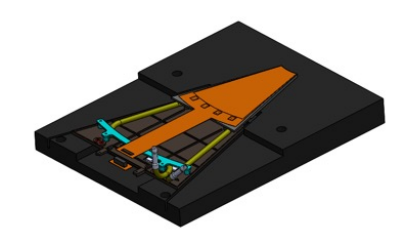}  
             \end{center}
         \caption{ The gantry picks up the Hybrid PBC and places it on the inner wedge. 
           \label{figs:module_robotic_inner_hybrid}}
          \end{figure}

    \item \textbf{Soldering the electronic components on the Hybrid PCB}: 
      All the electronic components are carefully soldered on the Hybrid PCB with a clean surface. Then, the flux is removed and the visual inspection is also performed. Finally, the open/short connection test is conducted to ensure the module is ready.  
 
\end{enumerate}

\subsection{Quality assurance}
\begin{figure}[!ht]
 	      \centering
 	      \includegraphics[scale=0.7]{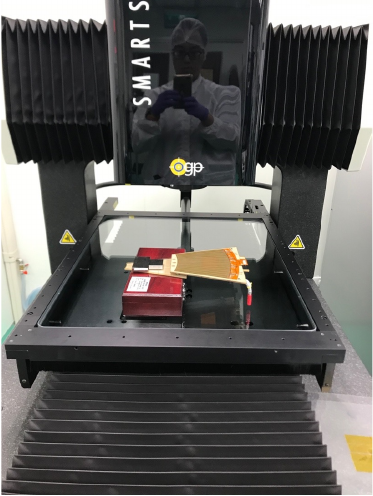}
         	\caption{The Optical Gauging Products (OGP) for STAR FST module assembly at TIDC.}
        	\label{figs:module_ogp}
        \end{figure}

After assembling the hybrid PCB and MS together, we perform similar measurements for flatness on the chip and sensor areas and positioning for the inner hybrid and outer hybrid (the distance from the center of the guide pins to reference points) using the Optical Gauging Products (OGP) at TIDC, as shown in Figure~\ref{figs:module_ogp}.

\begin{figure}[!ht]
 	      \centering
 	      \includegraphics[scale=0.5]{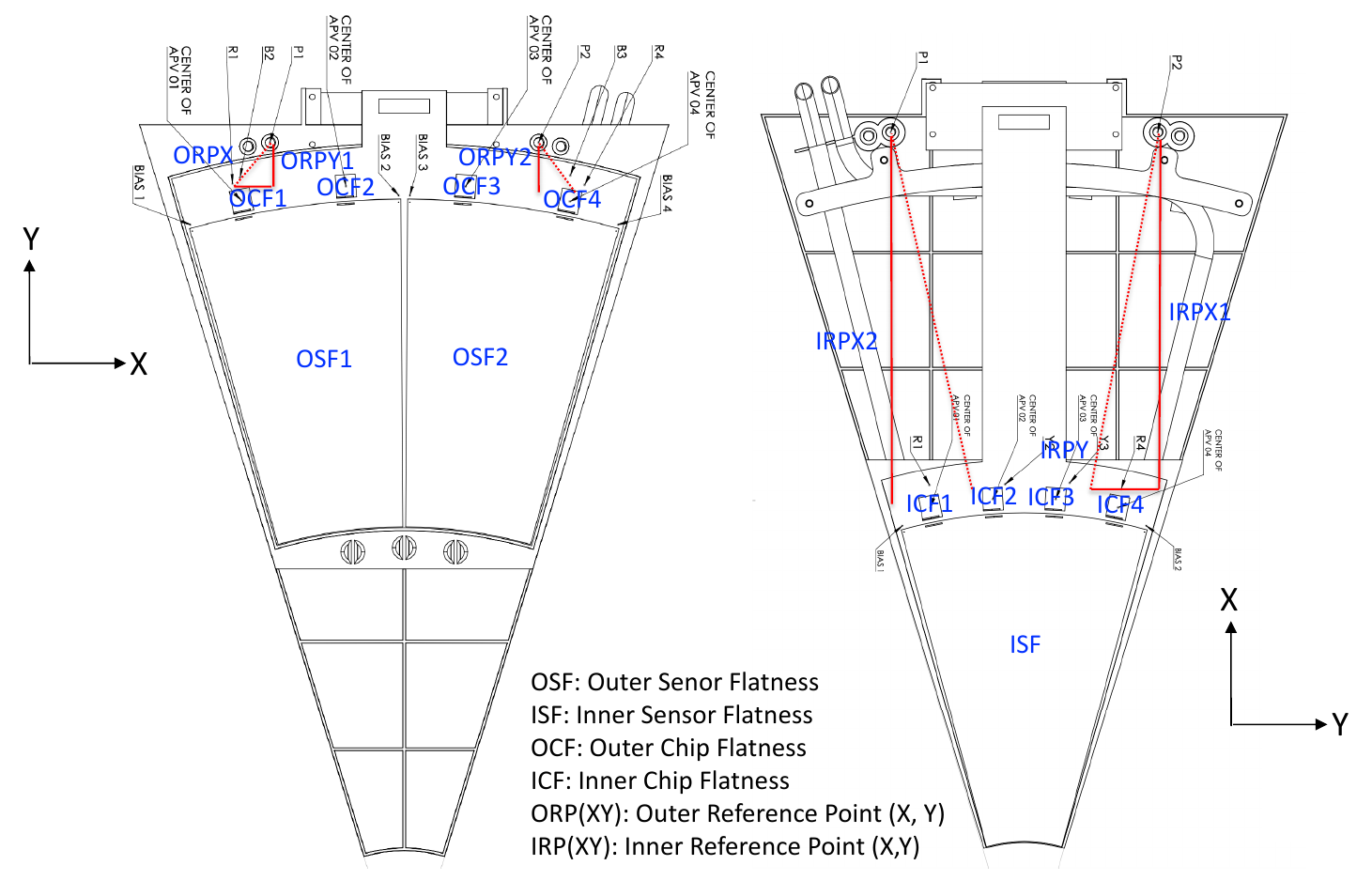}
         	\caption{The definition of STAR FST mechanical structure quality assurance.}
        	\label{figs:module_qa_definition}
        \end{figure}

There are several key variables for quality assurance, such as outer senor flatness (OSF), inner sensor flatness (ISF), outer chip flatness (OCF), inner chip Flatness (ICF), outer reference point (ORP) and inner reference point (IRP), as shown in Figure~\ref{figs:module_qa_definition}. 

Since the flatness on the sensor area is the key variable for the sensor assembly, a hard cut on the 250 $\mu$m on both OSF and ISF is applied. In addition, every module is graded by the criteria listed in Tab.~\ref{tabs:module_qa}. In the end, 73 modules are produced and there are 29 for Class A ($>$92), 31 for Class B (84 - 92), 5 for Class C ($<$ 84), and 8 for Class F (Failed).
\begin{table}
\begin{center}
\begin{tabular}{ |c|c|c|c|c|c| } 
 \hline
  Unit[$\mu$m]  & Weight & A (100 pts) & B (90 pts) & C (80 pts) & D (60 pts)\\ 
  \hline
 OSF $\times 2$ & 25\% & $<$ 171  & 171 – 205  & 205 – 270  & 270 – 351 \\ 
 \hline
 ISF & 25\% & $<$ 94  & 94 – 218  & 218 – 276  & 376 – 388  \\ 
 \hline
 OCF $\times 4$ & 10\% & $<$ 55  & 55 – 70 & 70 – 114  & 114 – 184 \\ 
  \hline
 ICF $\times 4$ & 10\% & $<$ 52  & 52 – 88  & 88 – 141  & 141 – 206  \\ 
 \hline
 ORP X & 5\% & $<$ 40  & 40 – 53  & 53 – 99  & 99 – 109  \\ 
 \hline
 ORP Y $\times 2$ & 10\% & $<$ 15  & 15 – 26 & 26 – 58  & 58 – 153 \\ 
  \hline
 IRP X $\times 2$ & 10\% & $<$ 22  & 22 – 44  & 44 – 100  & 100 – 349  \\ 
   \hline
 IRP Y & 5\% & $<$ 11  & 11 – 51 & 51 – 132  & 132 – 165  \\ 
 \hline
 \end{tabular}
 \caption{The grading matrix for STAR FST mechanical structure.}
        	\label{tabs:module_qa}
\end{center}
         \end{table}

\section{\label{sec:assembly}Assembly and Testing}

\subsection{Module Assembly at FNAL}
The last step of FST module assembly was done at Fermi National Accelerator Laboratory (FNAL), which includes the mounting of APV chips and silicon sensors, in addition to performing wire bonding and bond encapsulation. 
Each FST module has 8 APV chips (4 on inner hybrid and 4 on outer hybrid), 1 inner silicon sensor, and 2 outer silicon sensors. 
The assembly of APV chips and silicon sensors on hybrids was performed with precision assembly tooling constructed by the University of Illinois at Chicago (UIC) machine shop.
Due to the design of the FST module with inner and outer sensors on opposite sides of a module, the assembly was performed separately for the two sides. 
From January 2021 to Jun 2021, 48 modules were successfully assembled and the detailed assembly procedure can be found below.

\subsubsection{Mounting and Wire-bonding of  APV Chips on the Inner Hybrid}
\label{sec:assemblyInnerChip}
\begin{figure}[htb!]
\centering
\includegraphics[width=0.8\textwidth]{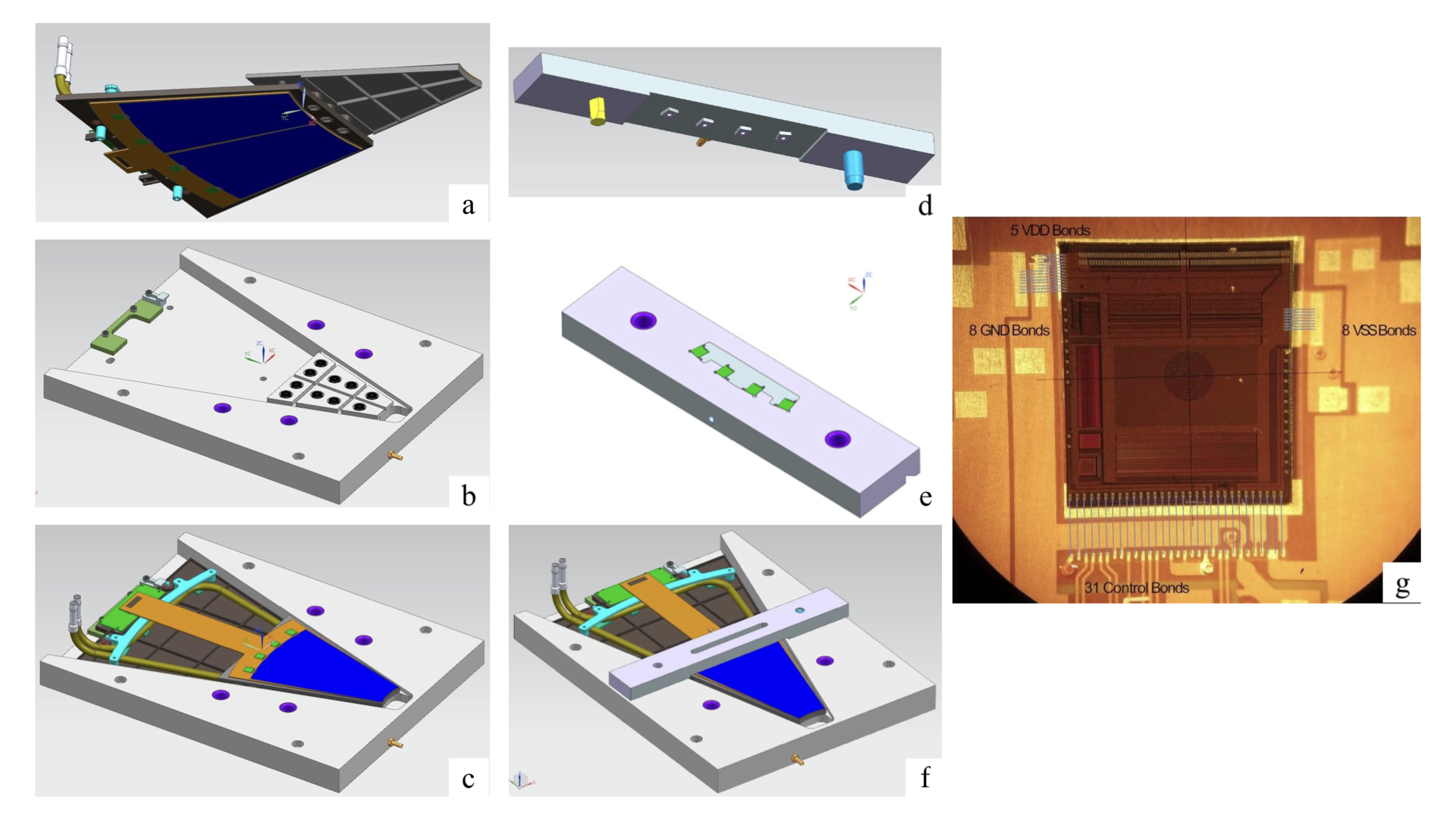}
\caption{The assembly procedure of APV chips on the inner hybrid.}
\label{figs:assemblyInnerChip}
\end{figure}

The mounting and wire bonding tooling and steps for APV chips are shown in Figure~\ref{figs:assemblyInnerChip}.
The bare FST module was first placed on the inner fixture with the inner sector faced up and fixed on the mechanical structure by the vacuum pads (Figure~\ref{figs:assemblyInnerChip} a to c). 
The position of a module with respect to the fixture was determined by the two alignment pins mounted on the FST module.
Next, the APV chips were placed into the inner stationing base and were secured on the base with a vacuum (shown in Figure~\ref{figs:assemblyInnerChip} e).
The APV chips were then picked up by the transfer plate and placed on the designed position on the inner hybrid through alignment holes on the inner fixture as shown in Figure~\ref{figs:assemblyInnerChip} f.
The chips were fixed on the inner hybrid by LOCTITE ABLESTIK 2902 silver epoxy and transferred to a dry box overnight.
The wire bonding and wire-bond pull tests were performed after the epoxy was fully dry. 
To perform the wire bonding, the humidity in the room was required to be 35-42\%. 
Figure~\ref{figs:assemblyInnerChip} g shows one APV chip after the wire-bonding procedure.
A readout test of the APV chips was performed for each module after mounting and wire bonding the APV chips.
The readout system at FNAL was the same as the readout system in the STAR experiment.

\subsubsection{Mounting  and Wire-bonding of  Inner Silicon Sensor on the Inner Hybrid}
\label{sec:assemblyInnerSensor}
\begin{figure}[htb!]
\centering
\includegraphics[width=0.8\textwidth]{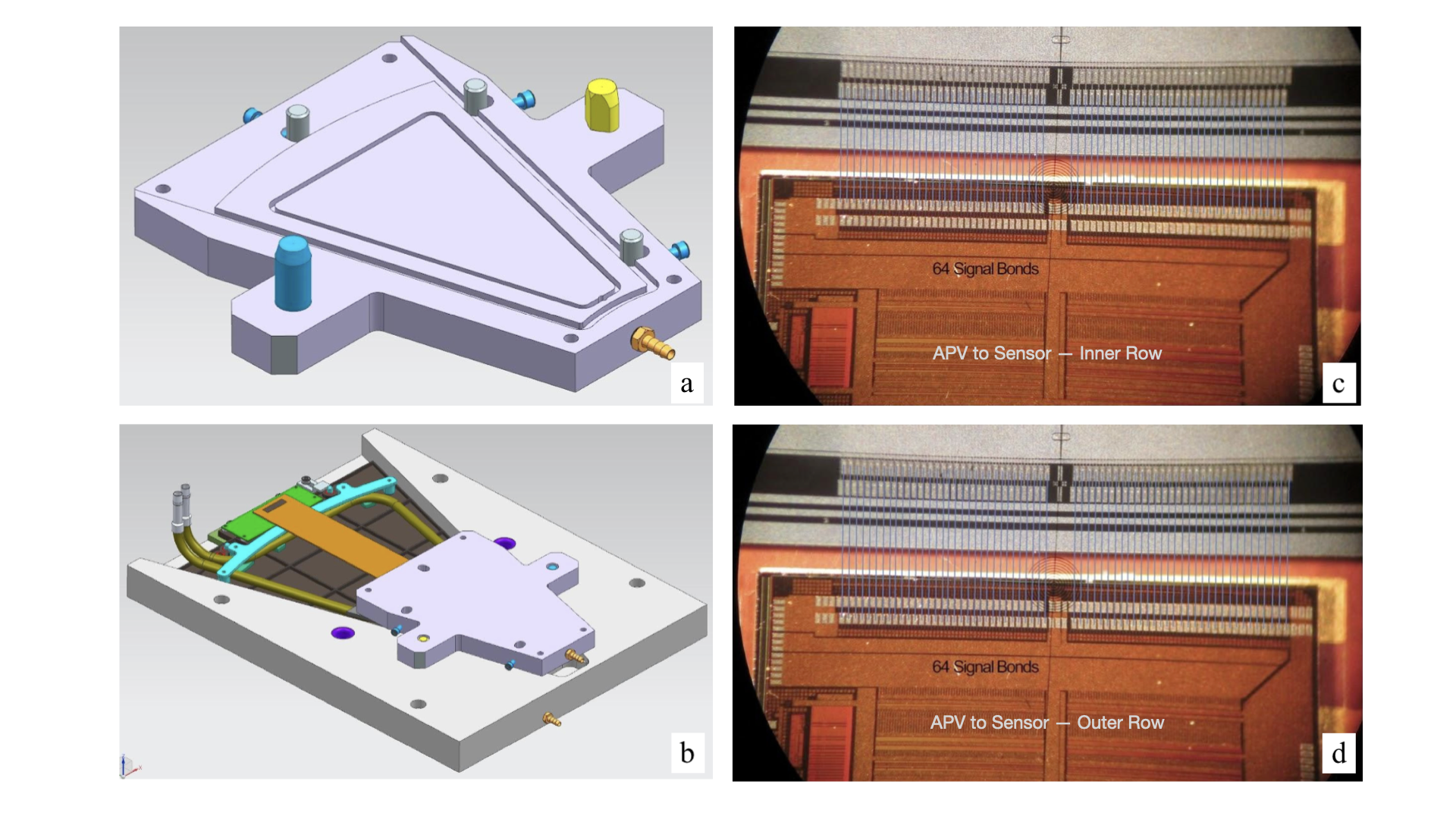}
\caption{The assembly procedure of inner silicon sensor onto the inner hybrid.}
\label{figs:assemblyInnerSensor}
\end{figure}

After a successful  readout test of APV chips, the FST modules were put back to the inner fixture and were held fixed with vacuum pads. 
An inner silicon sensor was then placed on the inner sensor placement fixture with the active surface down and and held in place by vacuum (Figure~\ref{figs:assemblyInnerSensor} a).
The TDR1100 and LOCTITE ABLESTIK 2902 epoxy were used to fix the inner silicon sensor on the inner hybrid.
After applying epoxy on the sensor area of inner hybrid, the inner sensor placement fixture with a sensor attached was flipped and then engaged with the inner assembly fixture. A weight was placed on the back of inner sensor placement fixture and the epoxy was allowed to cure overnight (Figure~\ref{figs:assemblyInnerSensor} b).
After the epoxy was fully cured, wire-bonding was performed (128 bonds per APV).
Figure~\ref{figs:assemblyInnerSensor} c and d show the APV chip after inner and outer rows of signal bonds were mounted on the chip and sensor.
A readout test was performed for each module after this step for the inner silicon sensor.

\subsubsection{Mounting and wire bonding APV chips and outer silicon sensors onto the outer hybrid}

\begin{figure}[htb!]
\centering
\includegraphics[width=0.8\textwidth]{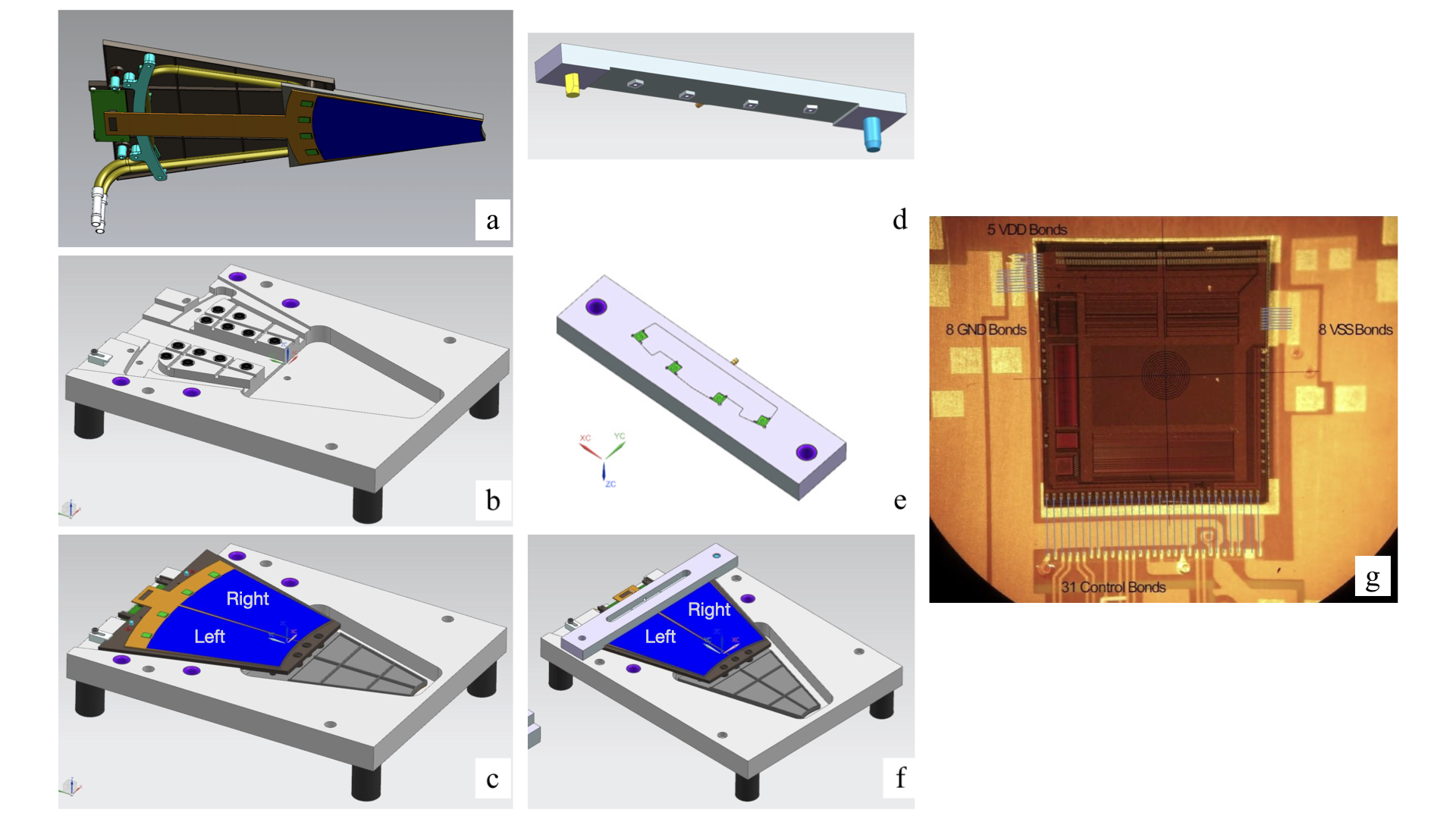}
\caption{The assembly procedure of APV chips on the outer hybrid.}
\label{figs:assemblyOuterChip}
\end{figure}

\begin{figure}[htb!]
\centering
\includegraphics[width=0.8\textwidth]{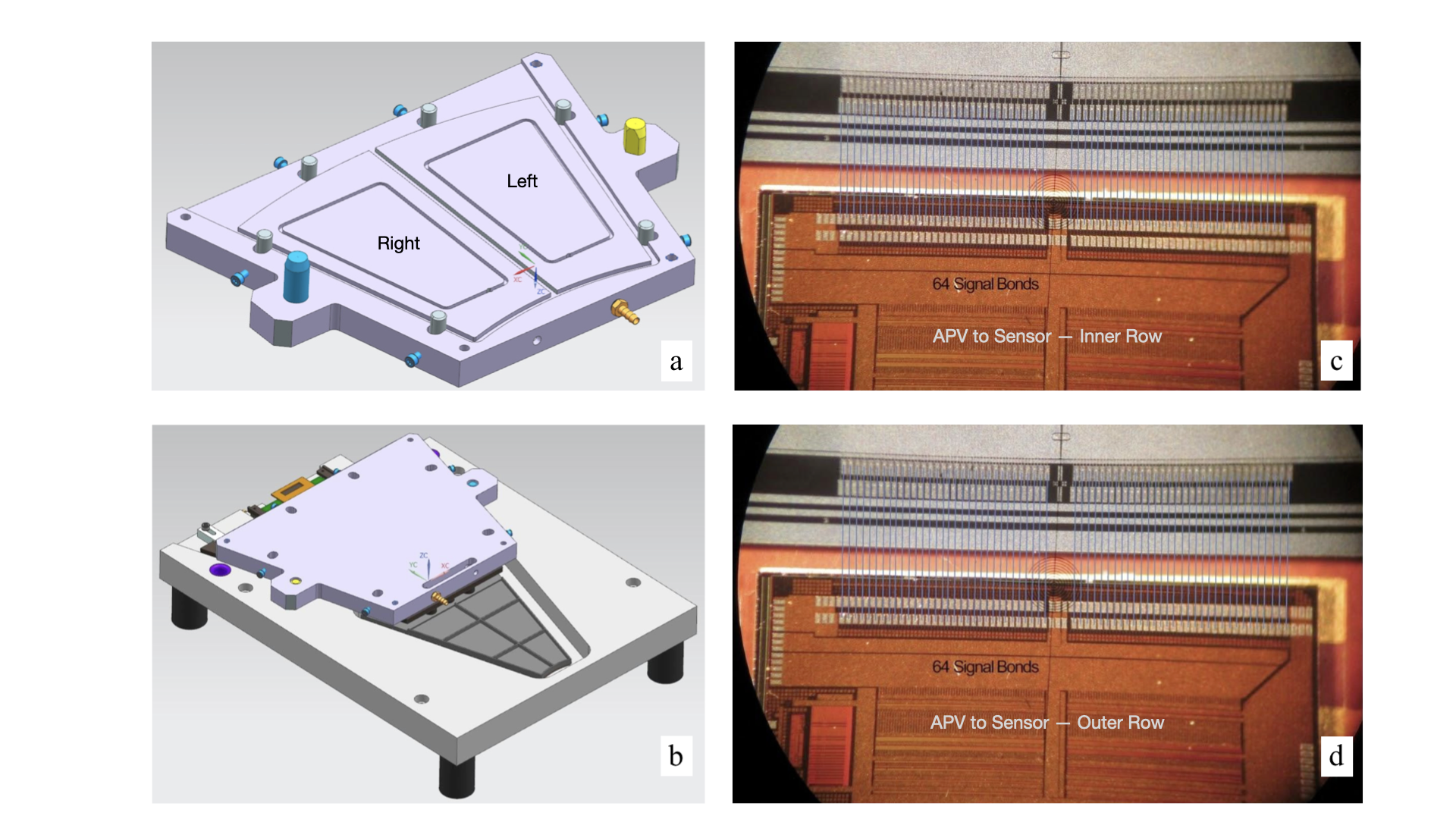}
\caption{The assembly procedure of outer silicon sensors on the outer hybrid.}
\label{figs:assemblyOuterSensor}
\end{figure}

The mounting and wire bonding procedures for APV chips and outer silicon sensors on the outer hybrid are very similar to the procedures for the inner sector presented in Sec.~\ref{sec:assemblyInnerChip} and ~\ref{sec:assemblyInnerSensor}.
The main difference is the layout design with two sensors and 2 APVs per sensor.
The detailed tooling and setup are shown in Figure~\ref{figs:assemblyOuterChip} and Figure~\ref{figs:assemblyOuterSensor}.

\begin{figure}[htb!]
\centering
\includegraphics[width=0.8\textwidth]{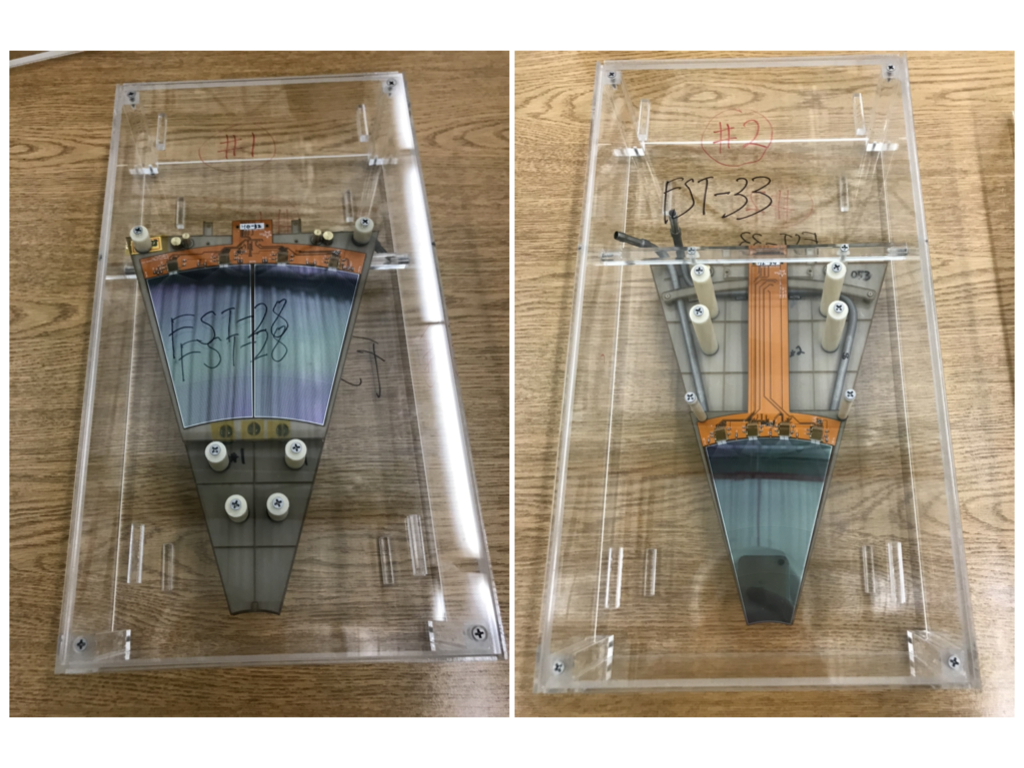}
\caption{Fully assembled FST modules in their storage boxes.}
\label{figs:assemblyModule}
\end{figure}

All the micro bonds and all sides of the APV chips were encapsulated with well-mixed resin (Dow Corning Sylgard 186).
A readout test was performed after the resin was fully cured.
Two fully assembled FST modules are shown in Figure~\ref{figs:assemblyModule}.

\subsection{Testing with Cosmic Rays at UIC}
\begin{figure}[htb!]
\centering
\includegraphics[width=0.8\textwidth]{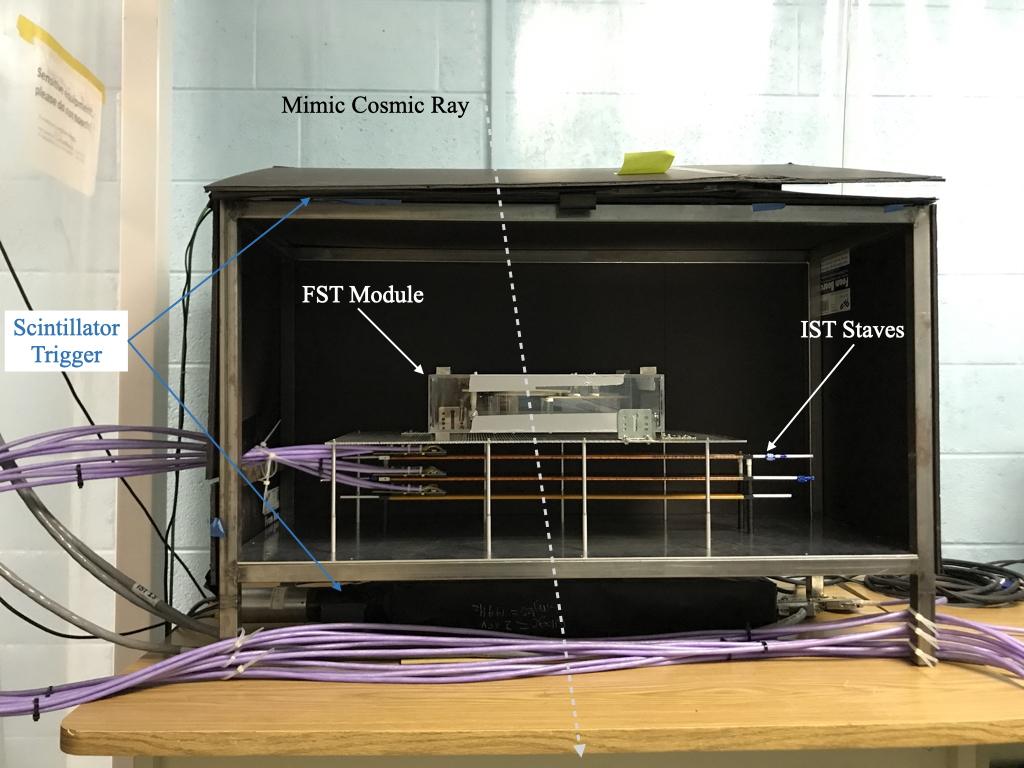}
\caption{The FST cosmic test stand at UIC. The white plastic rings are part of the holding box}
\label{figs:cosmicTestStand}
\end{figure}

To check the performance of the assembled FST modules, one of every eight assembled FST modules was brought to UIC for cosmic ray testing.
The cosmic test stand is shown in Figure~\ref{figs:cosmicTestStand}.
The assembled FST module was placed inside a dark box with three Intermediate Silicon Tracker (IST) staves underneath.
The IST staves were originally used in the STAR Heavy Flavor Tracker detector from 2014 to 2016 and were used as a proximate tracking device for the FST cosmic ray tests. Each IST stave consists of 36 APV25-S1 chips and 6 custom-designed silicon sensors (300$\mu$m silicon with 2 metallization layers). A detailed description of IST can be found in Ref \cite{Buck:2014mda}.
Two large scintillator detectors with PMT attached were located at the top and bottom of the dark box.
The coincidence signal of two scintillators was used as the trigger for cosmic rays with a trigger rate of about 0.5 Hz.

\subsubsection{Cosmic Test Result}
\begin{figure}[htb!]
\centering
\includegraphics[width=1.0\textwidth]{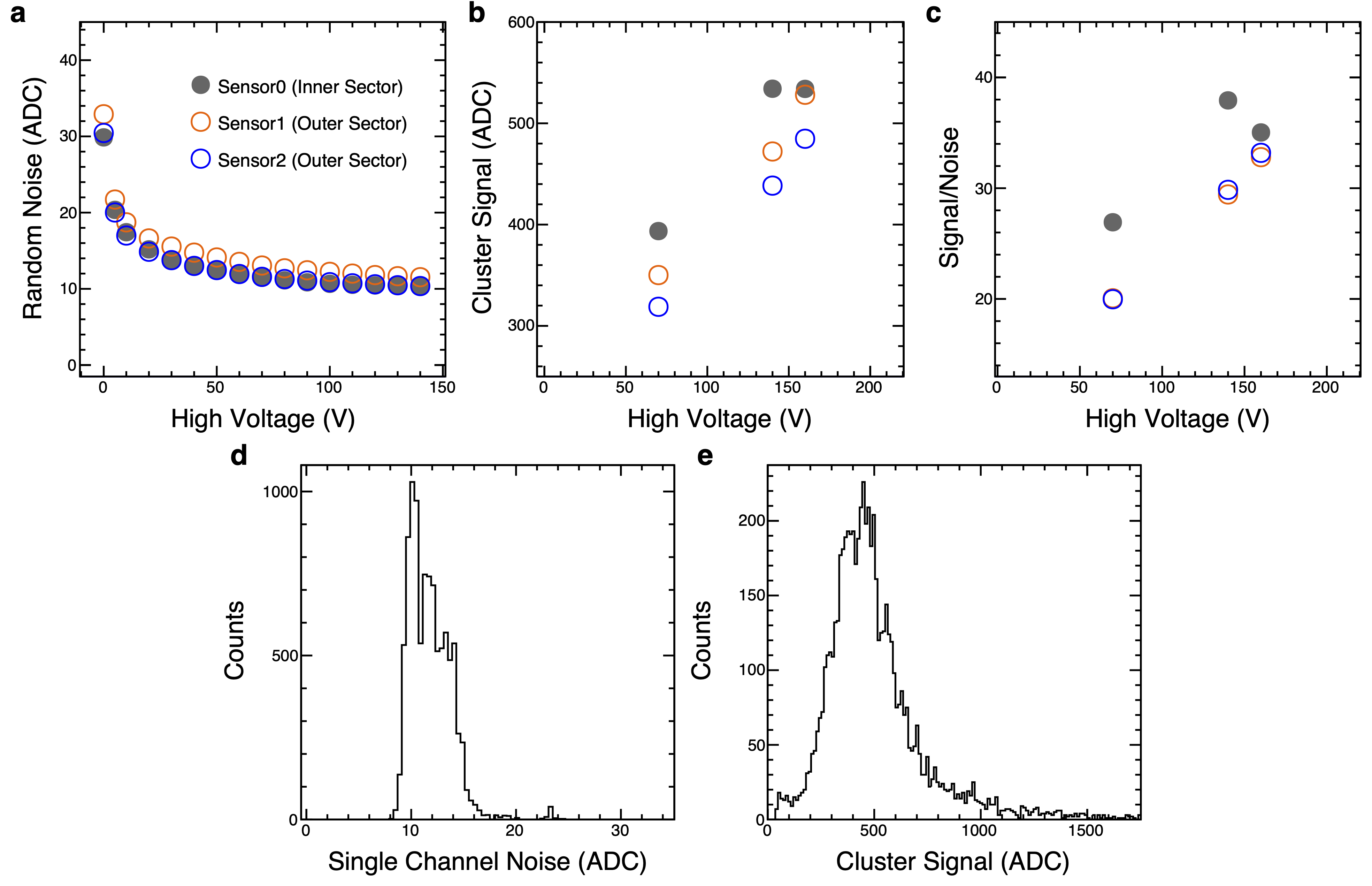}
\caption{Bias high voltage dependence of: a) average noise, b) cluster signal, and c) cluster signal to noise ratio. d) Distribution of noise for individual channels. e) Distribution of total cluster signals from cosmic ray tracks. }
\label{figs:cosmicSignal}
\end{figure}

To find the optimal operation voltage, a voltage scan was performed with cosmic rays detected by the FST modules.
Since the cosmic trigger rate was relatively low, only three voltage values (70V, 140V, and 160V) were tested for cosmic signal. 
The nominal operating voltages are 140 V for inner sensors and 160 V for outer sensors.
A voltage scan for noise was done for every 10 V, from 0 to 140 V, with an additional point at 5V.
The results are shown in the top panel of Figure~\ref{figs:cosmicSignal}.
The mean noise level at 140V was 10-15 adc counts, corresponding to about 600-900 e$^{-1}$, suggesting the input capacitance to APV ranges between 10-18 pF \cite{APV2}. The mean signal-to-noise ratio was 30-40 for different silicon sensors. The mean signal to noise on the eight individual radial strips from lowest radius to highest radius was 38, 32, 34, 36, 28, 31, 33, and 39.
The bottom panels of Figure~\ref{figs:cosmicSignal} present the distribution of noise and signal for all channels in one FST module.

\begin{figure}[htb!]
\centering
\includegraphics[width=0.75\textwidth]{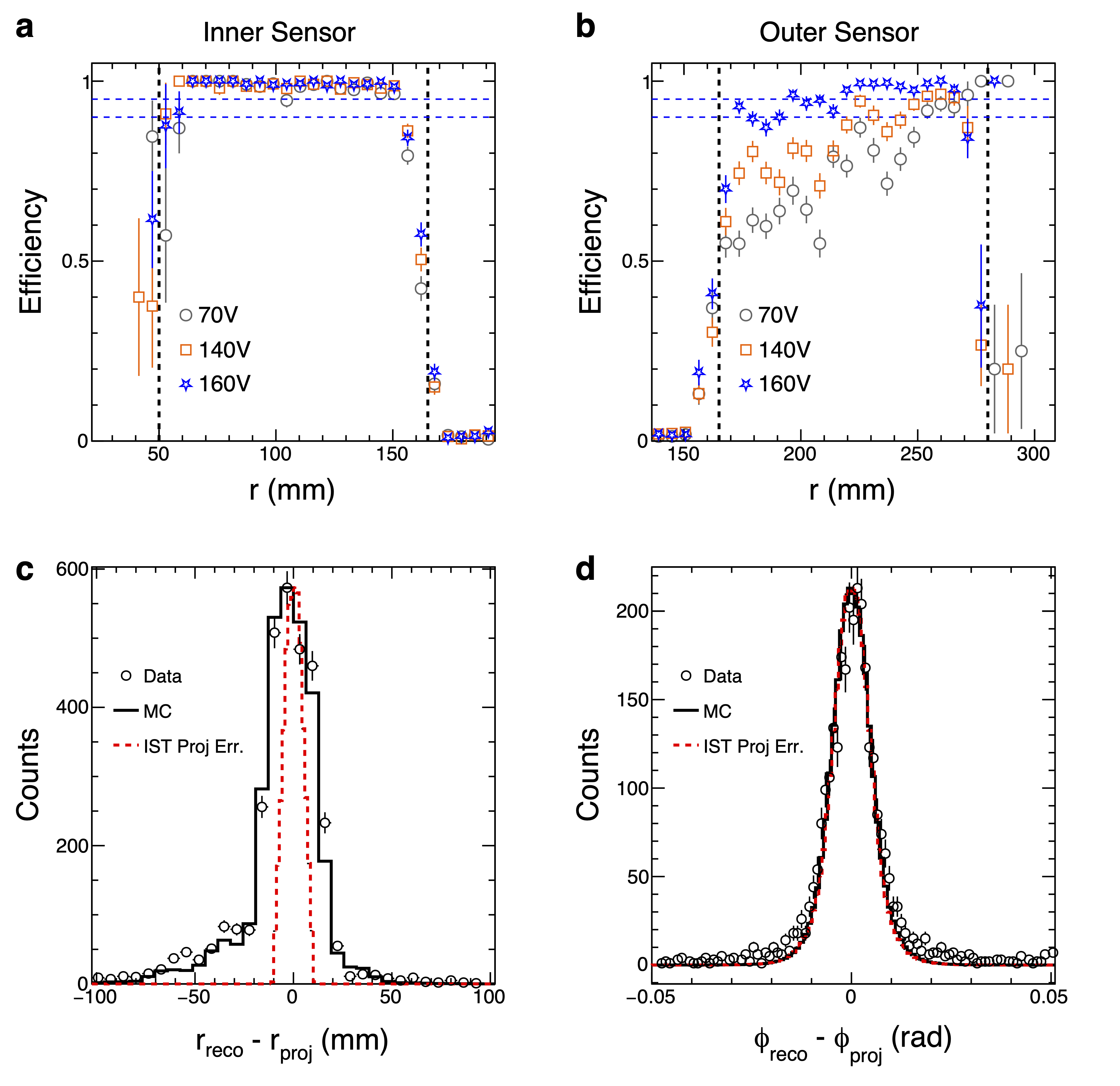}
\caption{a) Inner and b) outer sensor cluster detection efficiencies with respect to the radial position on the sensors. c) Radial and d) azimuthal angle residuals between reconstructed cluster positions and the expected position from the projected IST tracks. The black solid line is the expected distribution from a Monte Carlo simulation considering IST projection error (red dashed line).  }
\label{figs:cosmicEfficiency}
\end{figure}

The detection efficiency and position resolution were estimated with the help of IST.
For an ideal cosmic event, each IST stave will have a cosmic hit signal.
The incident position of the cosmic ray on the surface of an FST module is projected by the $\geq$2 hits obtained on 3 IST staves using a simple tracking prescription.
The detector efficiency and position resolution of the FST module are evaluated by comparing the readout signal of the FST module with the projected position from the IST staves.
The top panel of Figure~\ref{figs:cosmicEfficiency} presents the detector efficiency of inner and outer silicon sensors with different bias voltages applied.
The detection efficiencies of inner and outer silicon sensors are larger than 95\% and 90\% at 70V and above, which meets the physics requirement for the FST module.
The bottom panel of Figure~\ref{figs:cosmicEfficiency} shows the position resolution of an FST module in radial and azimuthal directions.
Due to the large projection error of IST, the extraction of position resolution is not straightforward; however, the cosmic data matches very well with a Monte Carlo simulation, which takes IST projection error into account.
In general, the assembled FST modules meet the key requirements of the physics program.

\section{\label{sec:electronics}Electronics}

As mentioned earlier the APV25-S1 chip~\cite{APV2,APV} is used for the FST front-end readout. It consists of 128 channels of charge-sensitive preamplifiers with pulse shapers, and 192 time-bin circular buffer analog memories. The analog memory is written at 37.53~MHz (four times the RHIC collision frequency), and memory locations are marked for readout at predetermined offsets from the STAR experiment trigger, which has a latency around 1.6~$\mu$s as received by the FST readout electronics. The APV latency register is typically set around 86; this, in combination with the fixed signal and cable delays and the other fixed latencies in the system results in the correct readout. In normal data taking for the FST, 3 memory locations, or "time-bins," are marked for readout. However, 9 or more timebins are used during commissioning and latency scans, to more clearly identify the correct triggered bunch crossing. 
The APV is operated in "20 MHz" mode, and "multi" mode rather than "de-convolution" mode, as defined in the users guide \cite{apvUserGuide}.

The FST utilizes a readout system for the APV chip which was developed for the STAR Forward GEM Tracker~\cite{APVreadoutSTAR} and IST~\cite{Buck:2014mda} (2014-2016) and further applied to the STAR GEM Monitor Tracker ~\cite{Ermakov:2016yzz} (from 2012). Details of the readout system are given in~\cite{APVreadoutSTAR} and only some highlights will be discussed here.

The readout system consists of two principal board types, the APV Readout Module (ARM) and the ARM Readout Controller (ARC). These are standard 6U~$\times$ 220~mm boards, housed in a common crate with the ISEG bias power supplies for the FST, mounted in a rack on the electronics platform adjacent to the STAR magnet where the majority of detector electronics systems and power supplies are situated. Each ARM handles 2 groups of up to 12 APV chips, with a common clock, trigger, and I$^2$C bus but separate APV data outputs. In the FST application, we use only 8 of the 12 possible APVs per group. Each ARC handles up to 6 ARM, thus up to 144 APV. The ARM's and APV are connected by a standard CPCI backplane although it is used with a simple custom protocol with LVCMOS registered transceivers. The ARC distributes the STAR clock and trigger signal to the ARMs, and routes configuration commands to them and to the APV chips through them. Additionally, when the ARMs have event data ready to send to DAQ, it gathers and packages the data and sends it, via the STAR-standard DDL link developed originally by CERN for ALICE~\cite{DDL}. The ARM distributes the clock and trigger-related APV write strobes (APV TRG signal), routes configuration commands to the APV through I$^2$C, and digitizes with 12~bit resolution the analog output readout data from the APV. The APV25 has a half-rate output mode which was intended for multiplexing two APV outputs locally at the front-end boards in CMS; we utilize this in order to reduce the bandwidth requirements on our long cable run. A complete frame of 128 channels (for a single time-bin) plus header from the APV chip takes 7.46~$\mu$s. The digitized data, including the APV digital header, is packaged and sent through the ARC to the STAR DAQ system. The ARM also provides isolated power to the APV chips; the I$^2$C is optically isolated at the ARM, and the differential analog inputs to the ARM although not isolated are designed for large common-mode rejection. All this facilitates operation with the readout mounted far (about 20~m) from the detector, which subjects the system to significant ground voltage noise, typically on the order of 0.5~V.

A set of patch panel boards are mounted about 1.5~m from the detector transitions between the small diameter custom cable used on the detector and the larger standard low-loss cable used for the long run to the readout system. The patch panel boards also remotely regulate the power fed from the ARM board to +/-1.25~V for the APV chips mounted on the detector hybrid.

\begin{figure}[htb!]
\centering
\includegraphics[width=0.37\textwidth]{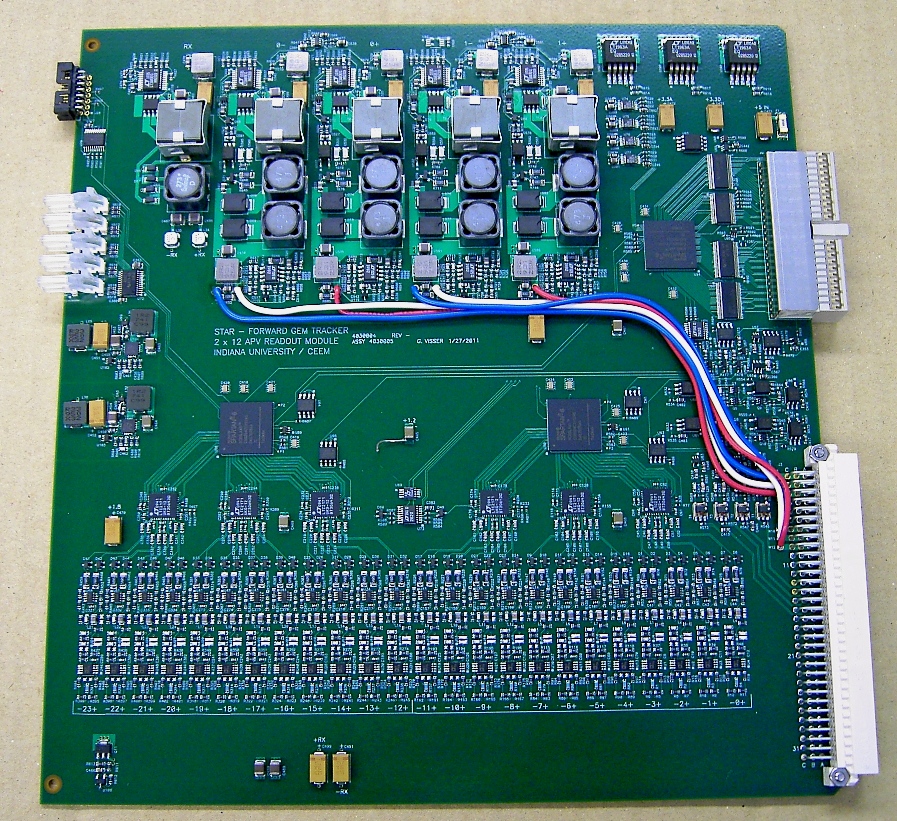}
\includegraphics[width=0.45\textwidth]{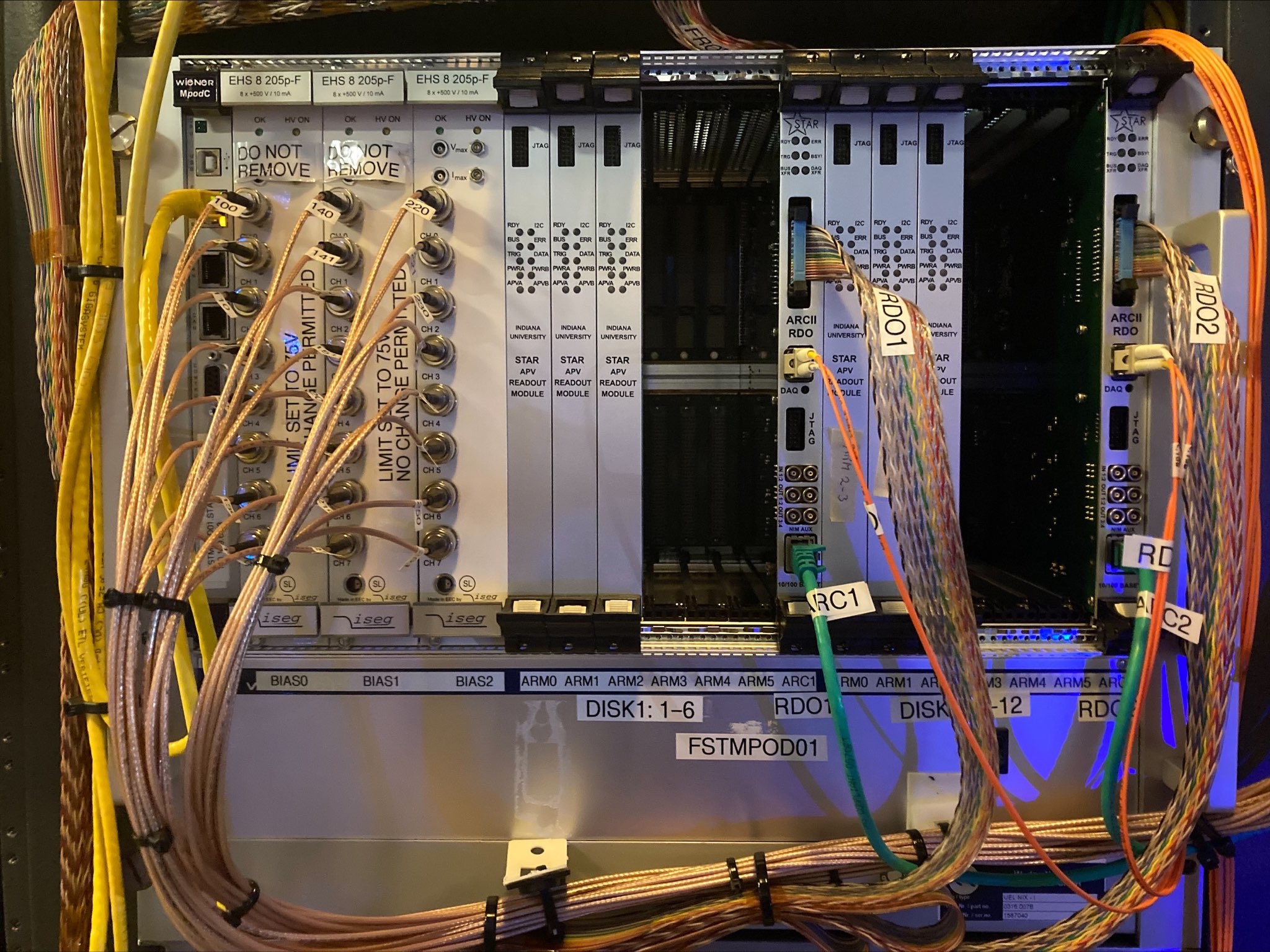}
\caption{APV Readout Module serving up to 24 APV chips.}
\label{figs:ARM}
\end{figure}

The original IST application utilized six ARC with six ARM each. All of this hardware was available to be repurposed to the FST, therefore we again utilize six ARC with now three ARM each. Splitting the load as much as possible in this way maximizes the trigger/readout rate capability.

\section{\label{sec:support}Support Structure and installation}

\begin{figure}[htbp]
\begin{minipage}{0.32\linewidth}
\centerline{\includegraphics[width=0.95\linewidth]{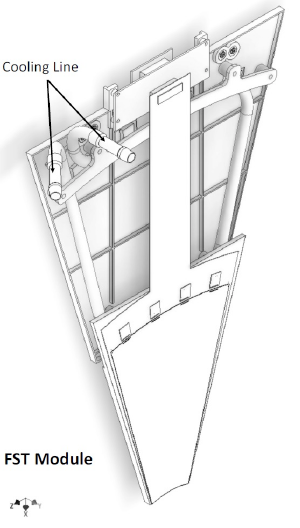}}
\end{minipage}
\begin{minipage}{0.32\linewidth}
\centerline{\includegraphics[width=0.95\linewidth]{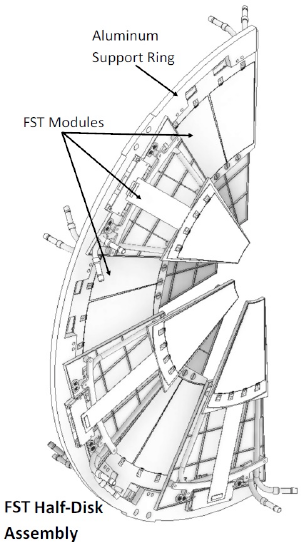}}
\end{minipage}
\begin{minipage}{0.32\linewidth}
\centerline{\includegraphics[width=0.95\linewidth]{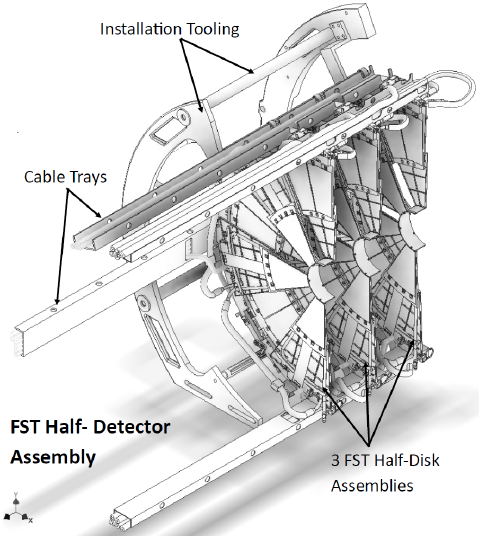}}
\end{minipage}
\caption[]{Left: FST module; Middle: FST modules assembled on half ring; Right: FST half detector assembly.}
\label{fig:FWD_module}
\end{figure}

48 FST detector modules as shown in the left panel of Figure \ref{fig:FWD_module} were delivered to BNL. Six modules were assembled on half rings forming half disk assemblies as shown in the middle panel of  Figure\ref{fig:FWD_module}. Three half rings were assembled using installation rails, tooling, cable trays, and cooling lines. Cables and cooling lines were installed in respective trays and FST half assembly was built as shown in the right panel of Figure \ref{fig:FWD_module}. During operation, each sector was cooled using a special cooling fluid (NOVEC 7200). NOVEC 7200 was used as it is non-conductive and is less prone to dripping or leaking (more details in Section \ref{sec:coolong}). Each half including installation tooling weighed around 100 lbs. FST detector was assembled in 2 halves as shown in Figure \ref{fig:FWD_module} because it had to be installed around the beampipe at STAR without disassembling the beampipe.  

\begin{figure}[htbp]
\begin{minipage}{0.49\linewidth}
\centerline{\includegraphics[width=0.95\linewidth]{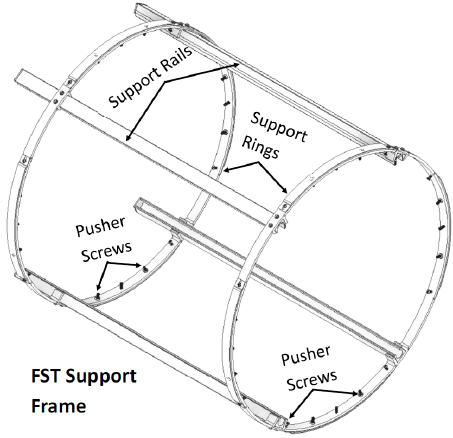}}
\end{minipage}
\begin{minipage}{0.49\linewidth}
\centerline{\includegraphics[width=0.95\linewidth]{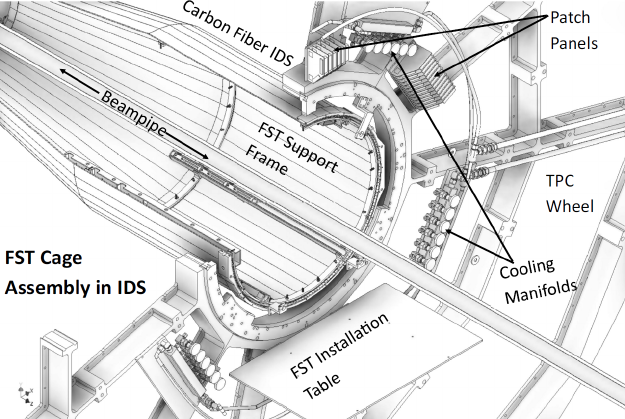}}
\end{minipage}
\caption[]{ Left: FST support frame; Right: Cage inside STAR magnet.}
\label{fig:FWD_cage}
\end{figure}

The FST detector’s final location inside the carbon fiber Inner Detector Support (IDS) structure of STAR as shown in the right panel of Figure~\ref{fig:FWD_cage} did not include any provision to install the FST detector. To facilitate the installation of the FST detector, a support frame as shown in the left panel of Figure \ref{fig:FWD_cage} was built and installed inside the IDS structure. It was centered around the beampipe using pusher screws. 

\begin{figure}[htbp]
\begin{minipage}{0.49\linewidth}
\centerline{\includegraphics[width=0.95\linewidth]{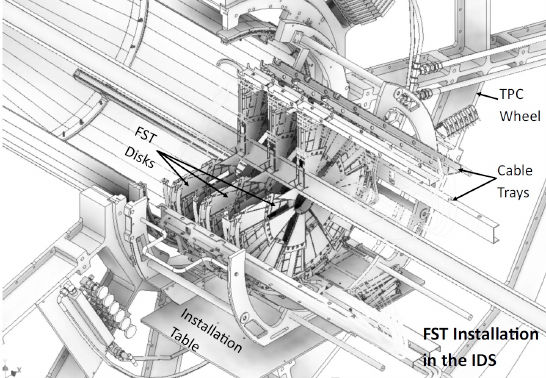}}
\end{minipage}
\begin{minipage}{0.49\linewidth}
\centerline{\includegraphics[width=0.95\linewidth]{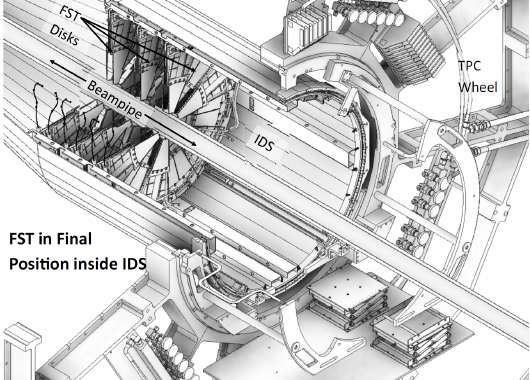}}
\end{minipage}
\caption[]{ Left: FST detector in assembly position on the installation table; Right: FST detector installed inside the STAR magnet.}
\label{fig:FWD_installation}
\end{figure}

Pre-existing threaded holes on the Time Projection Chamber (TPC) wheel and IDS support structure were used to mount the installation table, installation tooling as well as patch panels for cables and cooling manifolds. Two FST half assemblies were brought in using the crane. Both halves were placed and then assembled around the beam pipe using the installation table to form a full detector assembly as shown in the left panel of Figure \ref{fig:FWD_installation}. FST detector was then slid in on the installation rails to its final position inside the carbon fiber IDS as shown in the right panel of Figure \ref{fig:FWD_installation}. All the cooling lines were connected to the cooling manifolds and cables were connected to the patch panels. The installation table and tooling were removed from the TPC wheel after FST was installed in the final position.

\section{\label{sec:coolong}Cooling System}

\begin{figure}[htb]
\includegraphics[width=0.95\textwidth]{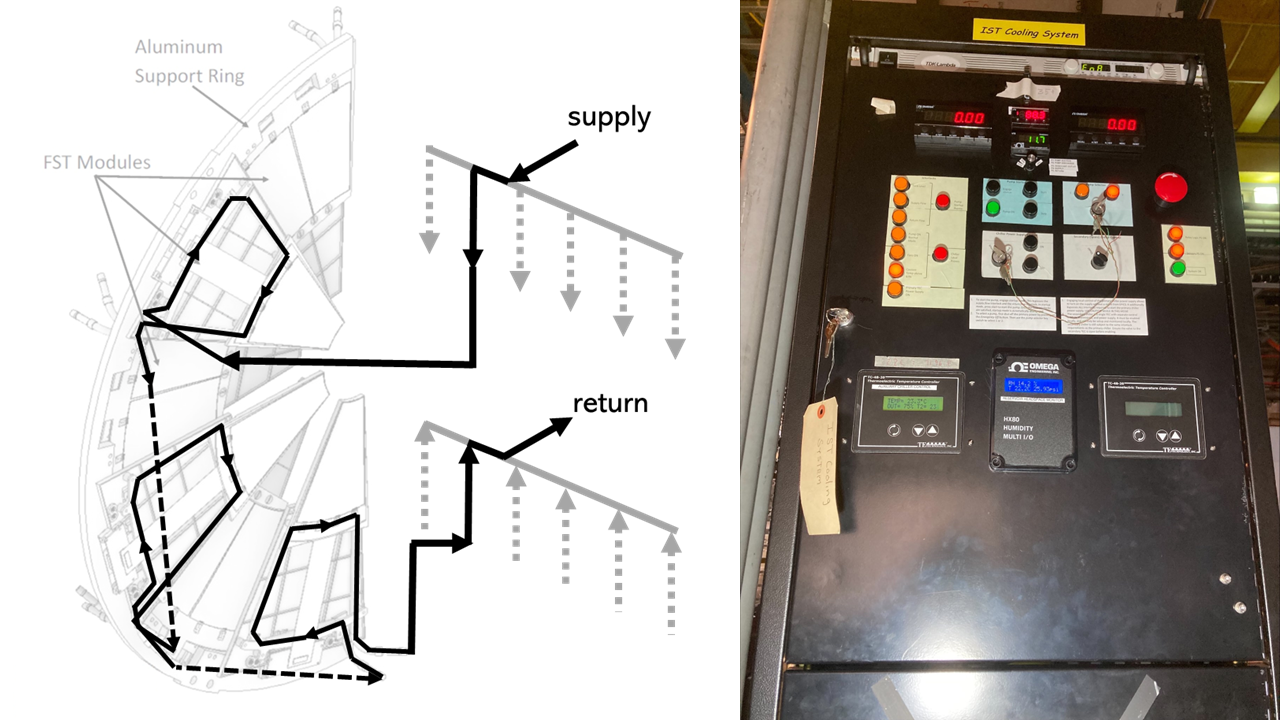}
\caption[]{Two pairs of supply and return manifolds are connected directly from the FST cooling rack. Each manifold has 6 loops. Left: Example of the FST cooling layout. Three FST modules are connected as one loop on one of the supply and return manifold pairs; Right: The front (operation) panel of the FST cooling rack, supports both in-person and remote control.}
\label{fig:fst-cooling-loop}
\end{figure}

The cooling system consists of five main components that maintain the FST temperature: 1) a reservoir with the cooling medium, 2) a pump for the fluid, 3) the heat exchangers, 4) the valves and filters, and 5) the power supplies. 
A 7-gallon ASME-certified pressure vessel is used as a fluid reservoir, which houses various sensors and maintains a pressure level of 25 psi. This reservoir contains the NOVEC 7200 engineered fluid used as the cooling liquid for this system. NOVEC 7200 has a density of 1420 kg$/$m$^3$, a vapor pressure of 13.1 kPa (at 20$^0$C), a boiling point of 76$^0$C, and a freezing point of -138 $^0$C. By design, NOVEC 7200 is nonconducting and volatile and therefore an ideal choice for our purposes. 

The NOVEC liquid is pumped by a magnetically-coupled rotary vane pump, produced by Fluid-O-Tech~\footnote{https://www.fluidotech.it/}. This pump has only one static seal which uses a Viton gasket. It includes an internal pressure bypass valve that is set to 100 psig. 
Two identical pumps are furnished for redundancy. Both pump units are wired and plumbed in parallel, while only one pump operates at a time. The active pump is manually alternated during FST operation. The speed of the pumps is set to 1100 rpm manually and left unchanged after commissioning. The heat from the circulating fluid is extracted using thermoelectric cold plates (TECs) connected to all-metal tube heat exchangers mounted on the cold side of the TECs (inside the rack). The hot side of the TECs is mounted to the outside of the rack, separated by a firewall. The heat from the hot side is rejected into the surrounding air by finned heat sinks and large fans. All the components of the heat exchangers are manufactured by TE Technology, Inc. 
During normal operation, three different TECs are operated in parallel, constituting the primary TEC system. In the event of failure or when the surroundings get extremely hot a secondary system can be engaged, also in parallel, consisting of a single TEC. 

The flow of the NOVEC liquid is regulated by bellow-type valves that utilize static seals. Such a design is less prone to leakage compared to packed seals in typical valves and has been recommended by the manufacturer (3M) of the NOVEC engineered fluid. All the wet parts of the valves are compatible with metals. The fluid circulation unit includes different types of filters that have several purposes. A 20-micron mesh screen filter is installed upstream of the pump to remove particulates from the cooling liquid. It is housed in a sanitary canister and is removable and cleanable. Water removal systems rely on the property of NOVEC as a part of the breather/reservoir system. In addition, granular-activated charcoal filters are employed to remove any dissolved contaminants such as plasticizers, solvents, and other organic impurities. For additional safety, the coolant liquid will be tested to have a water contamination level of 90 ppm or less (as recommended by 3M) before being poured into the reservoir. Karl Fischer titration of a sample of the NOVEC liquid may be necessary to ensure the water impurity is below 90 ppm.

Besides operating the pump and chiller by hand, the cooling system can be monitored and controlled by an input-output controller (IOC) software based on the MODBUS. 
The status of the pump and chiller (on or off), and the coolant temperature can be controlled remotely through a logic controller. All the necessary information like input/output flow speed and reservoir level are monitored simultaneously.

The temperature of the coolant is set at 22.2 $^0$C during the run. Because NOVEC is a volatile liquid, minor leaking of the NOVEC is inevitable.  The leak rate of the whole cooling system increased from 0.6\% to 0.9\% per day at the end of Run-22.

\section{\label{sec:operation}Operational Experience}

\begin{figure}[htb!]
\centering
\includegraphics[width=0.45\textwidth]{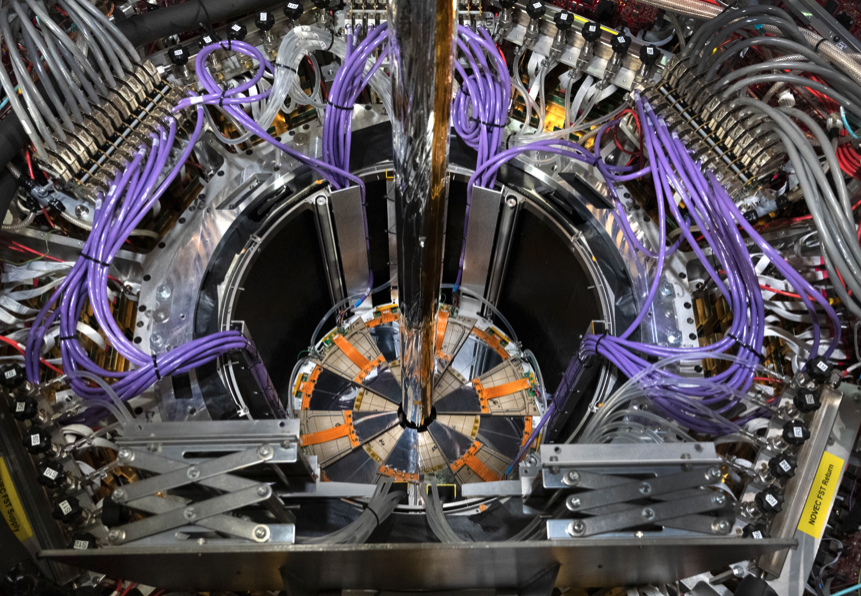}
\includegraphics[width=0.4\textwidth]{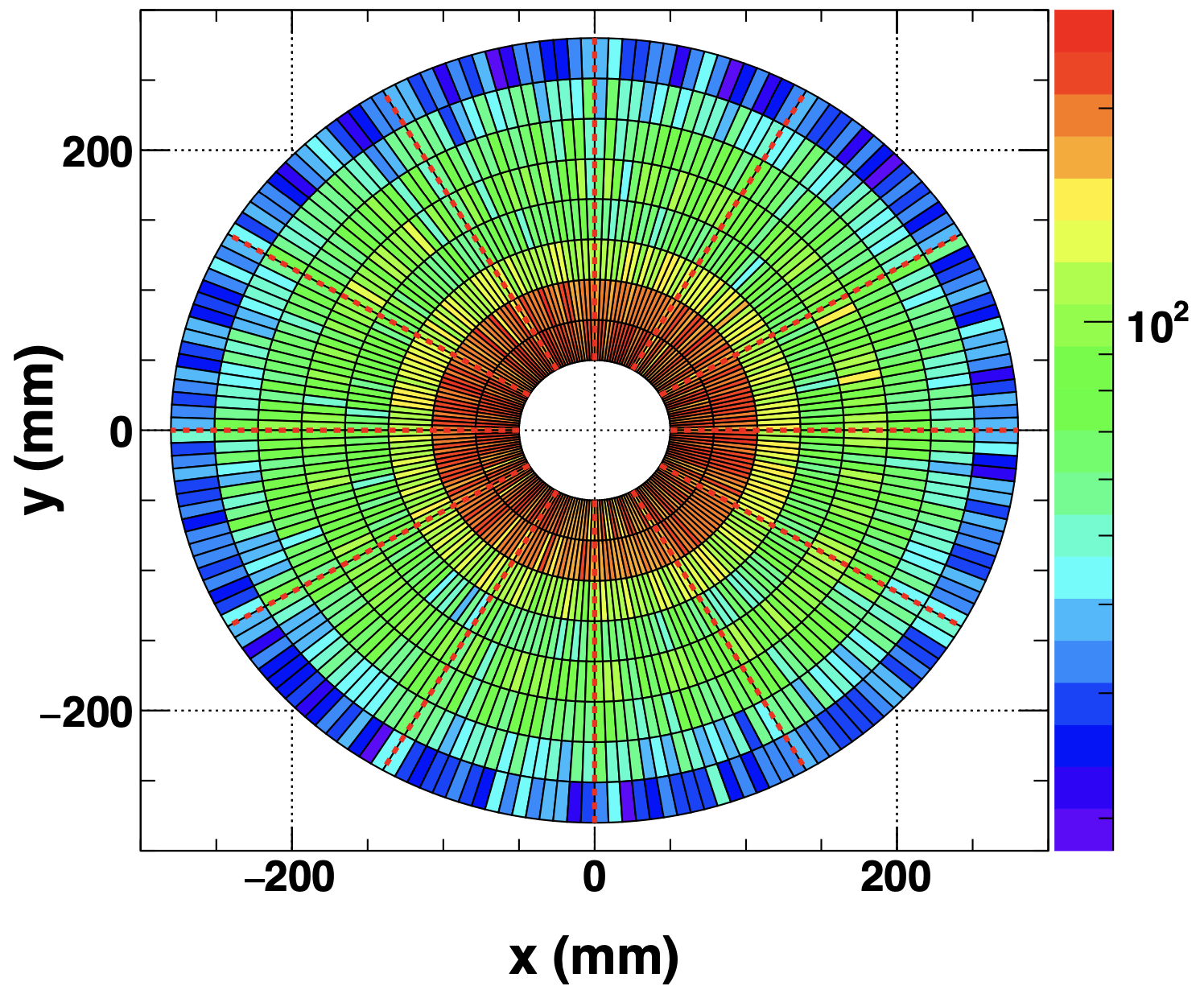}
\caption{Left: FST after installation; Right: cumulative event display for p+p \ 510 GeV collisions.}
\label{figs:operationFstRun22}
\end{figure}

The installation of FST was completed in August 2021 and the first $p+p$ 510 GeV collision data was recorded in December 2021. Figure~\ref{figs:operationFstRun22} (left) shows the FST installed in STAR and (right) the cumulative event display for one of the FST planes. The FST ran smoothly through the whole of Run-22 and Run-23. The detector operation was performed via slow control software, which requires minimal interaction by the shift crew.

To find the optimal operational high voltage, a voltage scan was performed with low luminosity runs in December  2021. It was decided to set the high voltage to 140V and 160V for inner and outer silicon sensors, respectively. The FST was running with 9 time bins initially for the detector commissioning and tuned to 3 time bins in December 2021 to increase the maximum DAQ rates of FST to 4.5kHz.

\begin{figure}[htb!]
\centering
\includegraphics[width=\textwidth]{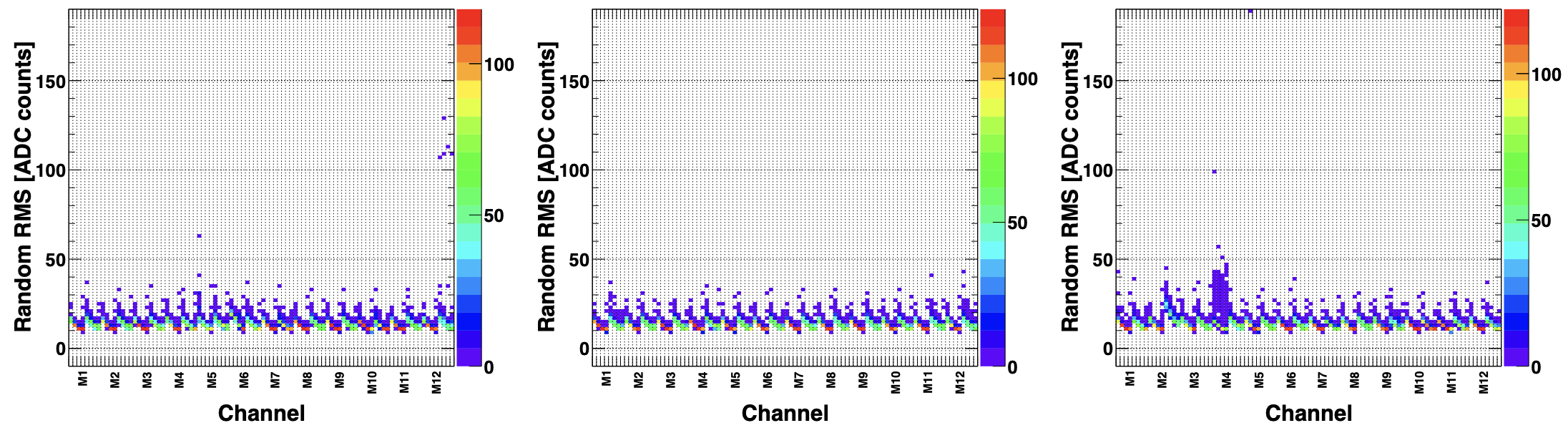}
\caption{Noise levels for FST silicon sensor channels. FST disks 1, 2, and 3 correspond to the left, center, and right figures, respectively.}
\label{figs:operationFstNoiseRun22}
\end{figure}

\begin{figure}[htb!]
\centering
\includegraphics[width=\textwidth]{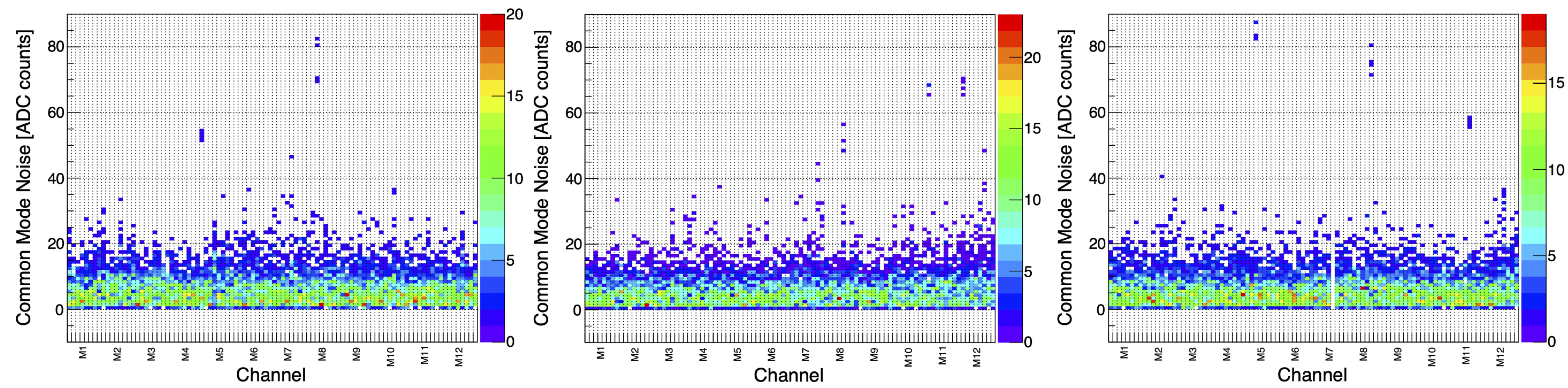}
\caption{Common mode noise for FST silicon sensor channels. FST disks 1, 2, and 3 correspond to the left, center, and right figures, respectively.}
\label{figs:operationFstCMNRun22}
\end{figure}

\begin{figure}[htb!]
\centering
\includegraphics[width=0.85\textwidth]{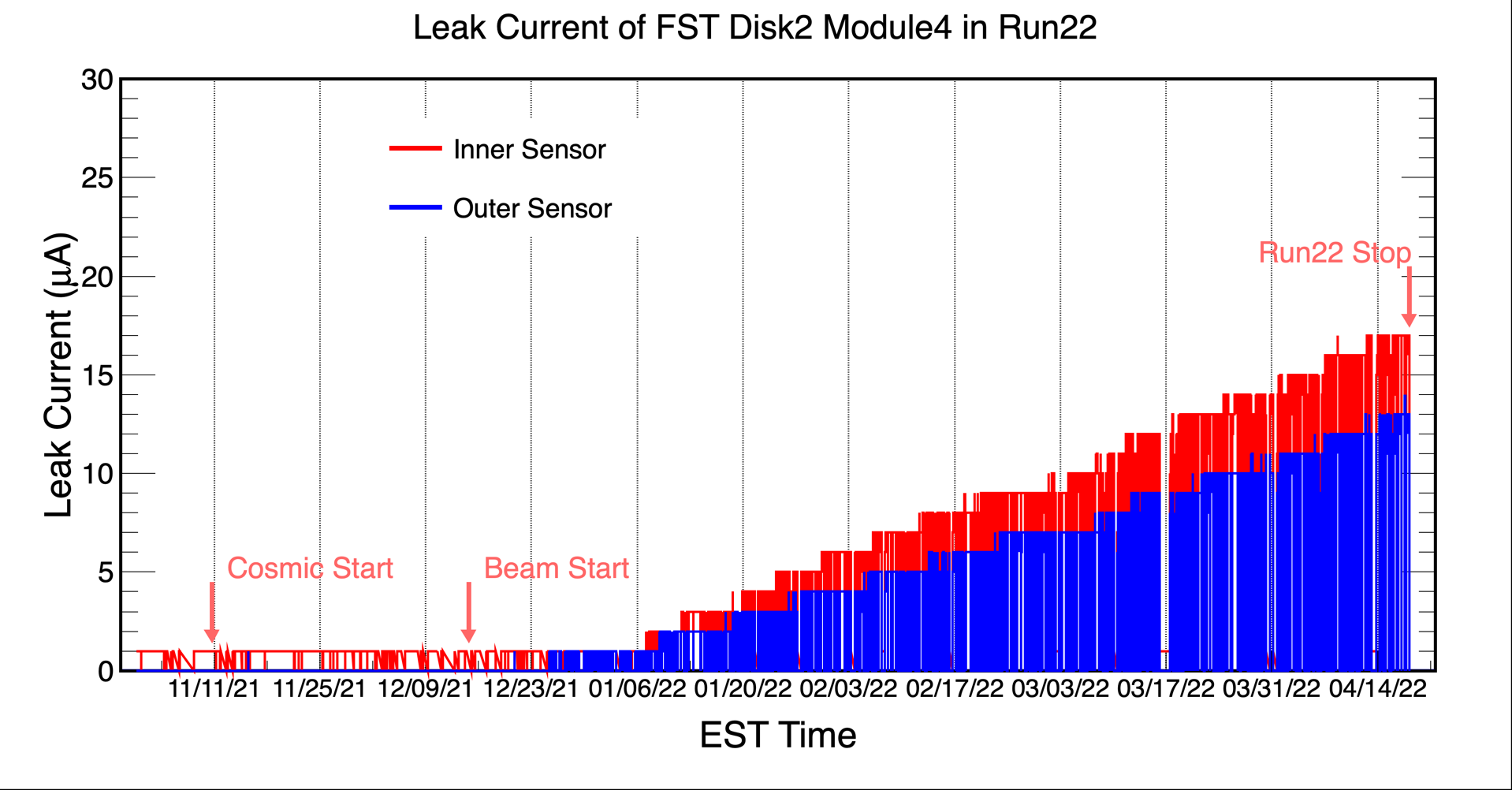}
\includegraphics[width=0.85\textwidth]{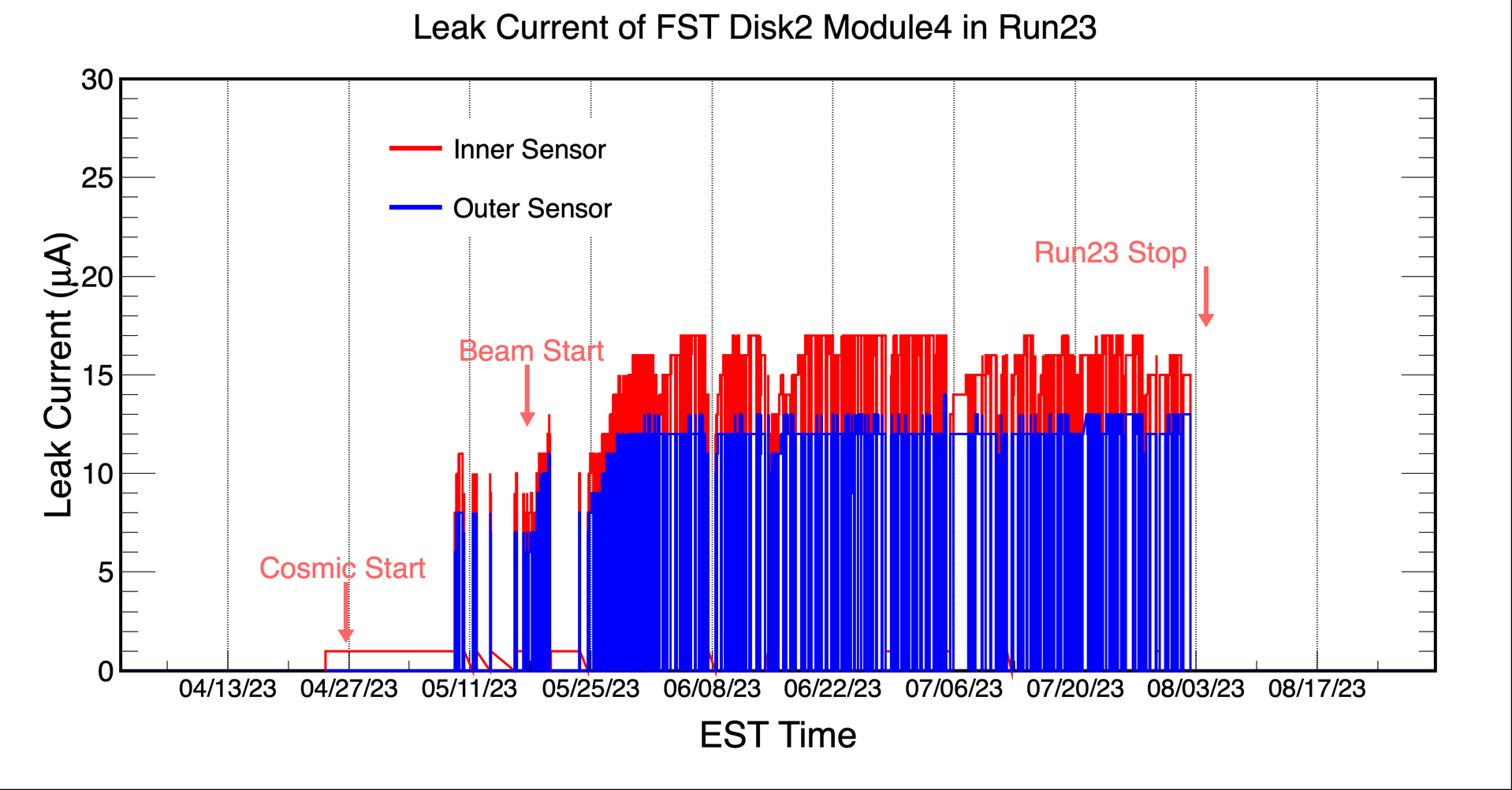}
\caption{The current of FST silicon sensors vs. time. \\}
\label{figs:operationFstCurerntRun22}
\end{figure}

The noise level of FST silicon sensors (shown in Figure~\ref{figs:operationFstNoiseRun22}) is 10 to 20 ADC counts depending on the position of the silicon strip and the average signal-to-noise ratio is about 25. The common mode noise for all channels is shown in figure ~\ref{figs:operationFstCMNRun22} and is generally below 15 ADC counts. Both random and common mode noise are extracted on an event-by-event basis. Due to irradiation damage, the leakage current of silicon sensors (shown in Figure~\ref{figs:operationFstCurerntRun22}) increased from 2 $\mu$A to around 13$\mu$A (inner silicon sensor) and 17$\mu$A (outer silicon sensor) after 4 months of p+p 510 GeV data taking due to irradiation damage. This increase is consistent with expectations. At the end of Run-23, the leakage currents of the sensors were roughly the same level as the end of Run-22, about 13$\mu$A and 17 $\mu$A for inner and outer silicon sensors as the beam luminosities were low in 2023. There were 2 inner sectors and 3 outer sectors operating at lower bias voltages due to abnormal bias current behavior. 

The FST readout chips are kept at room temperature by the cooling crate (same crate also used by Intermediate Silicon Tracker) running 3M NOVEC. The coolant tank was refilled every 6 weeks by experts. 

In January 2023, the entire FST was removed from the TPC cone and reinstalled. After this procedure, the leak rate was recorded at an acceptable value of 0.5\% per day. This reduction in leak rate was possibly due to more secure cooling line connections when the FST was reinstalled. 

\section{Summary}
Since its installation in 2021 and operation during the $p+p$ run in 2022 and the Au+Au collision run in 2023, the FST has been an integral part of forward rapidity physics program. The FST's compact design, equipped with segmented silicon strip sensors and AVP front-ends, supported by Kapton-based flexible hybrids to DAQ system, has adeptly handled the high occupancy and data acquisition rates up to 4.5 kHz. Even with the expected increase in leakage currents due to irradiation damage, the FST has proven to be resilient. The system's performance along with the NOVEC 7200 cooling system has been consistently optimal. The initial higher leak rate of the cooling system, attributed to potential air entrapment, has stabilized, affirming the system's robustness after post-installation adjustments. The successful data acquisition over the past two years bolsters our confidence in the ongoing $p+p$ run in 2024 and the forthcoming final RHIC run of Au+Au collisions in 2025. The FST is poised to function effectively until the projected end of the RHIC program in 2025.

\section{Acknowledgement}

The following people B.H., T.H., X.S., G.W., G.X., Z.Y., S.Z. and Z.Z. are supported under DOE contract DE-FG02-94ER40865. Y.H., R.S., P.T., and F.V. are supported under DOE contract DE-SC0012704. J.D.B is supported under DOE contract DE-SC0024189. J.D., Y.H., M.N., G.Y., and L.Y. are supported by the National Natural Science Foundation of China (NSFC) under Grant Nos. 11890710 and 11890713. The NCKU group (Y.C., H.H., Y.H., H.L., P.W., and Y.Y.) is supported by the Higher Education Sprout Project by the Ministry of Education at NCKU and the National Science and Technology Council of Taiwan. X.S. is also partially supported by the Strategic Priority Research Program of the Chinese Academy of Sciences under Grant No. XDB34030000. We extend our gratitude to the STAR Collaboration, the forward upgrade team, and the STAR Technical Group for their invaluable support and contributions. We express our appreciation to Jaroslav Adam, Elke Aschenauer, Timothy Carmada, William Christie, Tonko Ljubicic, Prashanth Shanmuganathan, Lijuan Ruan, Robert Soja, William Struble, David Tlusty and Zhangbu Xu for their guidance and support. Additionally, we thank Ronald Lipton and Gerrit Van Nieuwenhuizen for insightful discussions and thank Joseph Brandenburg for providing photo-realistic renders of the FST. We also recognize the support from the NCKU lab, the utilization of the FNAL Silicon Detector Lab facilities, and the contributions of the STAR technical support group, all of which have been pivotal to our progress.


 \bibliographystyle{elsarticle-num} 
 \bibliography{FST_NIM}





\end{document}